%% file: arXiv16.tex
\documentclass[letterpaper,twocolumn,10pt]{article}
\usepackage{usenix,endnotes}
\usepackage[table,xcdraw]{xcolor} 

\usepackage{graphicx} % Wrap figure around text
\usepackage[utf8]{inputenc}
\usepackage[table,xcdraw]{xcolor}
\usepackage{rotating}
\usepackage{pbox}
\usepackage{pifont}% http://ctan.org/pkg/pifont
\usepackage[table,xcdraw]{xcolor}
\usepackage{array}
\usepackage{multirow}
\newcommand{\cmark}{\ding{51}}%
\newcommand{\xmark}{\ding{55}}%
\usepackage{caption} 
\usepackage{comment}
\usepackage{xspace}
\usepackage{tikz}
\usepackage{color,soul}
\newcommand{\hlc}[2][yellow]{ {\sethlcolor{#1} \hl{#2}} }
\usepackage{xcolor}

\usepackage[compact]{titlesec} % reduce section spacing

\usepackage{textcomp,    % for \textlangle and \textrangle macros
            xspace}
\usepackage[fleqn]{amsmath}
\usepackage{listings}

\newcommand*\circled[1]{\tikz[baseline=(char.base)]{
            \node[shape=circle,draw,color=white,fill=black,inner sep=1pt](char){#1};}}

\newcommand{\system}{\textsf{Aware}\xspace}

\begin{document}

%don't want date printed
\date{}

%make title bold and 14 pt font (Latex default is non-bold, 16 pt)
\title{\Large \bf \system: Controlling App Access to I/O Devices on Mobile Platforms}
%\begin{comment}
%for single author (just remove % characters)
\author{
{\rm Giuseppe Petracca$^{\star}$}\\
gxp18@cse.psu.edu
\and
{\rm Ahmad Atamli$^{\ast}$}\\
atamli@cs.ox.ac.uk 
\and
{\rm Yuqiong Sun$^{\star}$}\\
yus138@cse.psu.edu
\and
{\rm Jens Grossklags$^{\star}$}\\
jensg@ist.psu.edu
\and
{\rm Trent Jaeger$^{\star}$}\\
tjaeger@cse.psu.edu\\ 
$^{\star}$The Pennsylvania State University\\
$^{\ast}$University of Oxford
% copy the following lines to add more authors
% \and
% {\rm Name}\\
%Name Institution
} % end author
%\end{commment}
\maketitle

% Use the following at camera-ready time to suppress page numbers.
% Comment it out when you first submit the paper for review.
\thispagestyle{empty}
%\AANote{title is a bit too long}

%\subsection*{Abstract}
%Paper submissions should be at most 13 typeset pages, excluding bibliography and well-marked appendices.
%Your Abstract Text Goes Here.  Just a few facts. Whet our appetites.

\input{abstract}

\section{Introduction}

\label{sect:intro}

Nowadays, mobile platforms are equipped with cameras, microphones and wide screens, which enable a variety of popular functions, ranging from audio/video recording to displaying information to users.  Many apps now utilize these functions to provide services that leverage on-board Input/Output (I/O) devices. For example, many apps now support voice and video messages, as well as photo/video shooting and editing. Even insurance or banking apps now utilize mobile platforms' cameras to collect sensitive information to expedite claim processing or check depositing \cite{esurance,pnc}. Apps able to record the screen content are also available for remote screen sharing or tutorial editing. 

However, uncontrolled access to on-board I/O devices can enable malicious apps, with access to these devices, to exfiltrate sensitive information.  Adversaries have built malware apps, referred to as {\em Remote Access Trojans} (RATs), that abuse authorized access to such devices to extract audio, video, screen content, and more, from mobile platforms, such as smartphones and tablets.  Malware in the wild both surreptitiously records data from a variety of mobile devices~\cite{lipovsky,rogers,npr}, but also performs directed attacks, such as constructing three-dimensional models of indoor environments~\cite{templeman} and extracting credit card numbers and PIN numbers~\cite{schlegel} from screenshots or keyboards' tones. Furthermore, uncontrolled accesses to on-board cameras and microphone can become sensitive if apps can stealthily take photos or videos and record audio by running a service in background\footnote{75\% of operations requiring permissions are performed when the screen is off or apps are running in background as services \cite{primal}}. 
Additionally, RAT apps can also use social-engineering techniques \cite{soc_eng} to hijack user-requested activities, such as showing a soft-button on screen that supposedly allows the user to take a picture when in reality the user action will trigger voice recording instead through the smartphone's microphone. 

Current defenses do not prevent malicious apps that happen to be granted access to I/O devices from exfiltrating sensitive data. Android, iOS, and Windows Phone OS all require users to authorize apps for access to I/O devices, such as the camera and microphone, at install time or at first use.  In many cases, users may assume that the use of such devices will be important, if not fundamental, for the effective operation of such apps.    Should the user authorize an app, the app can then choose when to use the device.  In addition, access to screen content is not even mediated by any of the mobile operating systems.  More restrictive security models, such as SE Android~\cite{smalley2013security}, cannot control apps access to such devices further, as they mainly protect the lower level Android system.  Some research systems aim to prevent unauthorized access to resources, such as these devices, by empowering apps to assist in the decision making~\cite{roesner2012user,nadkarni2014asm,backes2014android}.  However, since apps may be malicious, attacks cannot be prevented by this method.  Alternatively, researchers have explored auditing the use of such devices~\cite{xu2015semadroid} and providing visible indication of app behaviors~\cite{bianchi2015app}, but the former only detects attacks retroactively after the data has been exfiltrated, whereas user notification alone requires users to pay attention to each operation's security status, continuously, to avoid missing attacks. 

In this work, we propose the \system framework for authorizing app requests to perform operations using I/O devices, which binds app requests with user intentions to make all uses of I/O devices explicit.  To do this, we take the following steps.  First, app requests for operations using I/O devices are intercepted by the \system-enabled services to mediate all security-sensitive operations.  Second, \system makes each app request visible to the user, independently from the app, to enable users to express their intents without being spoofed.  Third, \system makes ongoing operations, targeting I/O devices, visible to users, so they may choose to terminate the operation.  As opposed to previous solutions, \system does not depend on apps to govern users, but rather \system links app requests and user input.  Also, \system maintains the security status of ongoing operations, so user actions are only necessary at operation initiation and completion, rather than requiring ongoing user monitoring  \cite{bianchi2015app}.

We have implemented \system on Android OS (android-6.0.1\_r5 version) and found it to perform effectively, by adding a maximum overhead of 4.79\% (minimum 2.19\%).  We have performed a user study, involving 74 human subject. Without \system, only 18\% of the study participants were able to identify attacks from tested RAT apps. \system systematically blocks all the attacks in absence of user consent and enabled study participants to identify 82\% of social-engineering attacks tested to hijack approved requests, including some more sophisticated forms of social engineering not yet present in available RATs.  This paper makes the following contributions:
\vspace{-2mm}

\begin{itemize}
\itemsep0em
\item We reverse-engineer two real-world RAT apps and two proof-of-concept RAT apps to systematically study and categorize the different techniques used by adversaries in mounting attacks targeting on-board I/O devices.
We identify five security properties that must be satisfied in order to ensure protection against stealthy operations from malicious apps targeting on-board I/O devices. 
%We also identify the run-time conditions sufficient to satisfy such security proprieties.
\item We propose \system, a security framework, that introduces defense mechanisms for enforcing these five security properties by mediating app requests to I/O devices and matching those requests to user consent, which blocks all unapproved requests and maintains the security status of ongoing requests to the user to enable prevention of social-engineering attacks.
\item We conduct an extensive user study involving 74 human subjects to evaluate: (1) users’ awareness of RAT attacks targeting I/O devices, (2) effectiveness of RAT attacks targeting on-board I/O devices, and (3) effectiveness of our proposed defense mechanisms in increasing users’ awareness and control over sensitive operations targeting I/O devices. 
\item We evaluate our approach on five RAT apps and eight widely-used apps, to show that it is possible to prevent against attacks from RAT apps without compromising functionality or introducing significant performance overhead.

\end{itemize}

\section{Problem Definition}

In  this section,  we  describe Remote Access Trojans (RATs) to demonstrate attacks that exploit use of I/O devices. We then examine the state-of-the-art in permission granting for mobile platforms to understand why such malware apps are capable of exploiting I/O devices on smartphones. We then outline the challenges for defenses capable of blocking such attacks.

\subsection{Remote Access Trojans (RATs)} \label{rats}

On smartphones, \textit{Remote Access Trojans} (RATs) are malicious apps users may be tricked into installing on their smartphones that aim to violate users' privacy and data confidentiality. RATs collect security-sensitive information through on-board I/O devices, such as cameras, microphones, and screen buffers using authorized app permissions to perform stealthy, malicious operations including taking photos and videos, recording audio, or capturing screenshots. 

Researchers have designed and developed mobile RATs to demonstrate limitations of current access control models adopted in mobile OSs. Examples include: \textit{PlaceRaider} \cite{templeman}, which uses the camera and other sensors to construct rich, three-dimensional, models of indoor environments; and \textit{Soundcomber} \cite{schlegel}, which exfiltrates sensitive data, such as credit card and PIN numbers, from both tones and speech-based interaction with phone menu systems.  Real-world RATs are also available online for purchase.  Two popular ones, widely discussed in security blogs and anti-virus companies, are:  \textit{Dendroid} \cite{rogers}, which takes photos using the phone’s camera, records audio and video, downloads existing photos, records calls, sends texts, etc. and \textit{Krysanec} \cite{lipovsky}, which takes photos, records audio through the microphone, locates victims via devices’ GPS, views victims’ contact list, and intercepts and sends text messages.

We reverse engineered and statically analyzed these RATs. The two real-world RATs were leaked online, whereas the two proof-of-concept RATs have been shared by researchers. From the analysis, we obtained details on how RATs work. We found out that: (1) \textit{all} analyzed RATs require a specific set of permissions to access on-board I/O devices, except when accessing the screen buffers; (2) \textit{all} of them run a background service that stealthily performs malicious operations to abuse permissions granted at install time or first use; (3) \textit{none} of their activities is shown on screen; (4) \textit{all} of them need access to the Internet to leak security-sensitive data collected; and (5) their stealthy operations are \textit{never} associated with any user interaction with the smartphone.

\subsection{Limits of Permission Granting}

Mobile OSes currently support two default mechanisms to grant apps permission to access on-board I/O devices.  

First, in Android OS, users grant apps permission to access I/O devices at \textit{install time}. Apps receiving permission at install time can then access I/O devices \textit{at any time}, without further user approval, so users are unable to track \textit{how} and \textit{when} sensitive on-board I/O devices are accessed by apps at runtime. Table \ref{permission_analysis} summarizes the permission sets required by the analyzed RAT apps to perform stealthy operations. We found out that: (1) \textit{all} permissions used by RAT apps to perform stealthy operations are classified as \textit{dangerous} by the official Android OS documentation \cite{AndroidDoc}; (2) users are \textit{never} notified about accesses to security-sensitive on-board I/O devices\footnote{Users could notice the Wi-Fi or cellular network icon on the phone screen status bar, but they do not know what app is responsible for the network traffic, and what data is flowing out through the network.}%and sensors\footnote{The user could notice that the GPS receiver is being used, by observing the GPS icons on the phone screen status bar, but the user does not have enough information about what app is reading the GPS coordinates.}
at runtime, via on-screen prompts, or notifications; and (3) the \textit{same} sets of permissions are used by well-known benign apps downloaded by millions of users worldwide. 
%As an example, photography apps (e.g., Instagram) require the CAMERA, WRITE\_EXTERNAL\_STORAGE, and INTERNET permissions to function as expected. 
%As a consequence, RAT apps can mask themselves behind more innocuous and benign apps.  
We performed an extensive analysis of permission sets used by apps by randomly selecting 74 apps from third-party app stores \cite{MobileApkWorld,ApksFree} and 329 apps from the official Google Play \cite{GooglePlay}. The results cause concern: 83.89\% of apps from the official Google Play store could potentially take stealthy screenshots, whereas 25.68\% of apps from third-party app stores could potentially take stealthy photos (complete analysis results are reported in Appendix \ref{perm_ineff}).

\begin{table}[t!]
\scriptsize

\centering
\caption{Android Permissions used by RAT apps}
\label{permission_analysis}
\setlength{\tabcolsep}{.2em} % REDUCE HORIZONTAL PADDING

\begin{tabular}{cccccccc}
\cline{2-4}
\multicolumn{1}{l|}{}           
& \multicolumn{1}{c|}{\cellcolor[gray]{0.7}} 
& \multicolumn{1}{c|}{\cellcolor[gray]{0.7}}
& \multicolumn{1}{c|}{\cellcolor[gray]{0.7}} \\

\multicolumn{1}{l|}{}           
& \multicolumn{1}{c|}{\multirow{-2}{*}{ \cellcolor[gray]{0.7}\begin{tabular}[c]{@{}c@{}}Permissions required \\to perform Operation\end{tabular}} }
& \multicolumn{1}{c|}{\multirow{-2}{*}{\cellcolor[gray]{0.7}\begin{tabular}[c]{@{}c@{}}Protection\\ Levels\end{tabular}}}
& \multicolumn{1}{c|}{\multirow{-2}{*}{\cellcolor[gray]{0.7}\begin{tabular}[c]{@{}c@{}}User\\ Notified\end{tabular}}} \\ \cline{2-4} 

%\multicolumn{4}{|c|}{{\cellcolor[gray]{0.6}On-Board I/O Devices}} \\ \hline 

\multicolumn{1}{|c|}{\cellcolor[gray]{0.8}\begin{tabular}[c]{@{}c@{}}Stealthy Photo\end{tabular} }        & \multicolumn{1}{c|}{ \begin{tabular}[c]{@{}c@{}} {\tiny CAMERA} \\ {\tiny WRITE}\_{\tiny EXTERNAL}\_{\tiny STORAGE}\end{tabular}}                   & \multicolumn{1}{c|}{Dangerous}   & \multicolumn{1}{c|}{No}           \\ \hline
\multicolumn{1}{|c|}{\cellcolor[gray]{0.8}\begin{tabular}[c]{@{}c@{}}Stealthy  Video\end{tabular}}        & \multicolumn{1}{c|}{\begin{tabular}[c]{@{}c@{}}{\tiny RECORD}\_{\tiny AUDIO, CAMERA} \\ {\tiny WRITE}\_{\tiny EXTERNAL}\_ {\tiny STORAGE}\end{tabular}}                   & \multicolumn{1}{c|}{Dangerous}   & \multicolumn{1}{c|}{No}           \\ \hline
\multicolumn{1}{|c|}{\cellcolor[gray]{0.8}\begin{tabular}[c]{@{}c@{}}Stealthy  Audio\end{tabular}}        & \multicolumn{1}{c|}{\begin{tabular}[c]{@{}c@{}}{\tiny RECORD}\_{\tiny AUDIO}  \\ {\tiny WRITE}\_{\tiny EXTERNAL}\_{\tiny STORAGE}\end{tabular}}                   & \multicolumn{1}{c|}{Dangerous}   & \multicolumn{1}{c|}{No}           \\ \hline
\multicolumn{1}{|c|}{\cellcolor[gray]{0.8}\begin{tabular}[c]{@{}c@{}}Stealthy  Screenshot\end{tabular}}        & \multicolumn{1}{c|}{\begin{tabular}[c]{@{}c@{}}{\tiny WRITE}\_{\tiny EXTERNAL}\_{\tiny STORAGE}\end{tabular}}                   & \multicolumn{1}{c|}{Dangerous}  & \multicolumn{1}{c|}{No}            \\ \hline
\multicolumn{1}{|c|}{\cellcolor[gray]{0.8}\begin{tabular}[c]{@{}c@{}}Remote Data Transfer\end{tabular}}        & \multicolumn{1}{c|}{{\tiny INTERNET}}                   & \multicolumn{1}{c|}{Dangerous}   & \multicolumn{1}{c|}{No\textsuperscript{1}}           \\ \hline

\end{tabular}
\end{table}

Second, starting from Android OS 6.0 (Marshmallow) and in other mobile operating systems, such as Apple iOS and Windows Phone OS, users are prompted with access requests \textit{the first time} apps request access to I/O devices. Researchers have shown that, while these prompts attempt to verify users' intent, in practice, they create an excessive burden on users, which leads to users ignoring these prompts eventually\footnote{Prompts are disruptive and cause excessive fatigue, conditioning user to simply accept any prompt query, resulting in undermining the usefulness of the prompts \cite{felt2012,yee2004aligning, primal}.}. In addition, these mobile operating systems allow users to manage permission grants at runtime by accessing a per app or per I/O device permission control panel. This feature allows for better flexibility in permission granting, since it is now possible to revoke, at runtime, permissions granted to apps at install time or first use. 

\begin{figure}[h]
\centering
\includegraphics[width=80mm]{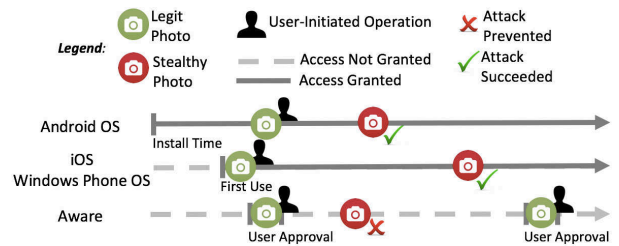}
\caption{Effect of the lack of binding between users' interaction and operations performed by running apps}
\label{fig:timeline}
\end{figure}

Unfortunately, neither of these mechanisms ensure that sensitive operations targeting on-board I/O devices are performed \textit{only} in response to users' interaction with running apps, which \textit{must unmistakeably bind the user's consent with specific security-sensitive operations targeting an I/O devices}. In the absence of such binding,  malicious apps are free to leverage I/O devices, even while running as background services, once they can trick users into granting them permissions as shown in Figure~\ref{fig:timeline}.

\vspace{-8px}

%\subsection{Challenges in Protecting On-Board I/O Devices from Exploits}
\subsection{Challenges in Preventing RAT Exploits}

Researchers have identified several attacks targeting on-board I/O devices  \cite{templeman,schlegel}, and proposed various defense mechanisms \cite{roesner2012user,roesner2014world,smalley2013security,AndroidEnh}. Unfortunately, significant attacks are still capable of circumventing proposed defenses. Furthermore, most anti-malware tools available for smartphones are not able to identify apps behaving as RATs, especially if not advertised and commercialized as spying tools on the Web\footnote{We have tested the 15 most popular Android anti-malware tools, complete results are reported in Appendix \ref{anti}.}. Similarly, current static analysis tools \cite{VirusTotal,chen2015finding,tam2015copperdroid,bouncer} designed to analyze apps' source code to  identify malice are often not able to find stealthy operations targeting I/O devices\footnote{We have tested 2 static and 2 dynamic analysis tools currently adopted by researchers and the general mobile app community.  Complete results are reported in Appendix \ref{stealthystalker}.}.

On one hand, several attempts have been made, in the past, to include users or surrounding environments in the access control decision mechanism.  Unfortunately, User-Driver Access Control (UDAC) \cite{roesner2012user} is subject to social-engineering attacks \cite{bianchi2015app}, whereas World-Driven Access Control (WDAC) \cite{roesner2014world} limits the objects that may be recorded, but does not enable users to control when apps use I/O devices for recording.  However, solutions to prevent social-engineering attacks \cite{bianchi2015app}  currently require users to pay attention to the security status of an operation throughout its execution, rather than just at the beginning and end.  On the other hand, researchers have also proposed methods to augment contemporary access control models based on the Android Permission mechanism \cite{AndroidDoc, felt2012} and SE Android \cite{smalley2013security,AndroidEnh}.  Android Security Modules (ASM) \cite{nadkarni2014asm} enable apps to assist in security decision-making, but unfortunately the apps performing operations are the adversary we must control;  whereas Android Security Framework (ASF) \cite{backes2014android}, which operates at the kernel level, does not have the necessary information about higher level events required to detect security-sensitive operations performed by processes running apps.  Additionally, solutions that leverage hardware support for app isolation, such as Samsung Knox \cite{knox}, would prevent apps in one domain from stealing sensitive data from apps in other domains, but RAT apps can still perform stealthy operations within their own domain.

Despite the various efforts above, several open challenges remain to be addressed:

%while designing mechanisms to regulate access by apps to on-board I/O devices in smartphones:

\begin{itemize}
\itemsep0em

% app uses I/O device anytime
\item Once apps obtain permission, at \textit{install time} or at \textit{first use}, they may stealthily access sensitive I/O devices \textit{at any time}.

% app uses wrong input to 
\item Mobile operating systems lack a method to {\em connect user interactions with security-sensitive operations} targeting on-board I/O devices for controlling access to such operations.
\begin{comment}
may not be \textit{always} aware and in control of security-sensitive operations targeting on-board I/O devices that could impact their privacy and data confidentiality\footnote{Researchers have argued that access control permissions need to be contextual, otherwise users do not know \textit{when} and \textit{how} permissions are used at runtime by installed apps \cite{primal}. }.
\end{comment}

% tricks users into providing
\item Apps may use \textit{social-engineering} techniques \cite{soc_eng,anderson2015supporting,bianchi2015app} to hijacking user-intended activities and trick users in authorizing undesired operations.

%to trick users about \textit{how} and \textit{when} I/O devices are accessed, resulting in users authorizing undesired operations.
\end{itemize}

In addition, any effectitve solution to these problems must only require user input and attention consistent with normal application use. 

\section{Background}

In this paper, we focus our attention on Android OS due to its open-source nature, availability, and popularity \cite{market}. Similar considerations hold true for other mobile operating systems, such as Apple iOS and Windows Phone OS. In the following subsections, we provide background information useful to understand the mechanisms proposed in this paper.

\subsection{I/O Device Management in Android} \label{io_background}

We briefly describe how processes obtain access to on-board I/O devices in Android OS. For performance and security reasons, only the Linux kernel has direct access to on-board I/O device drivers. The Hardware Abstraction Layer (HAL) implements an interface that allows system services (privileged Android processes) to indirectly access on-board I/O devices via well-defined APIs. SE Linux for Android guarantees that only system services can access on-board I/O devices at runtime. Thus, apps must communicate with system services, through the Binder mechanism \cite{AndroidDoc}, to request execution of specific operations  targeting on-board I/O devices. The system services designed to handle requests from apps would then execute the operations on behalf of processes running apps, if and only if, the necessary permissions have been granted to the requesting apps by the user or operating system. Permissions are validated by the \emph{Package Manager}, part of Android OS, each time apps request to perform operations targeting I/O devices. For instance, for operations targeting the microphone, the \emph{Package Manager} checks the apps \texttt{AndroidManifest.xml} files to verify if the RECORD\_AUDIO permission has been granted to the requesting app. If the required permissions are not granted to apps, the \emph{Package Manager} fires up a security exception to communicate the operation abortion. Android services do not require a permission check for apps to access the screen buffer.

\subsection{Android UI Graphical Elements}

\begin{figure} [t!]
\centering
\includegraphics[width=80mm]{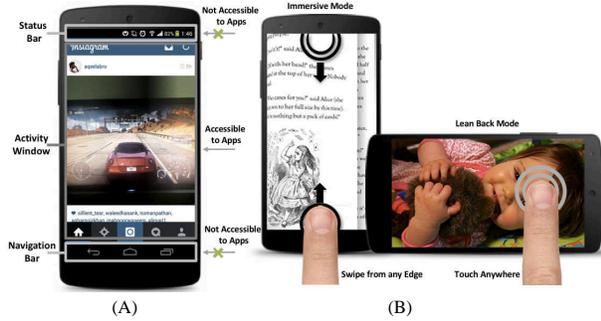}
\scriptsize{(A) \hspace{32mm} (B)  \hspace{10mm}}
\vspace{-3mm}
\caption{Android User Interface Graphical Elements and Gestures to bring back the Navigation and Status Bars into view when apps are in full screen mode}
\label{fig:screen}
\end{figure}

The Android User Interface is composed by three main graphical elements, as depicted in Figure \ref{fig:screen} (A). Different manufacturers may different layouts, but all distributions of the Android OS have the same three main elements.  

The \textit{Status Bar} shows the device's state, such as battery level and network connectivity. The \textit{Navigation Bar} includes three navigation buttons to interact with all currently running apps and the home screen.
The \textit{Activity Window} is the only portion of the screen that processes running apps can draw graphical elements on, such as \textit{Activities} and \textit{Views} inside activities. All activities created by apps are drawn in the system-managed \textit{Activity Window}. Activities are organized in a stack managed by the \textit{ActivityManager} system service, the only process allowed to manage these three main graphical elements. The only activity shown to the user is the one on top of the stack, even though, previously displayed activities might be visible if the new activity on top is only partially covering the \textit{Activity Window}.  

The \textit{Activity Window} becomes the only graphical element visible on screen when apps go in \textit{Full Screen} mode. Starting from version 4.4, the Android OS offers apps two approaches to go full screen: \textit{Lean Back}\footnote{Used when users won't be interacting heavily with the screen while consuming content, like while watching a video.} and \textit{Immersive}\footnote{Mainly intended for apps in which the user will be heavily interacting with the full screen as part of the primary experience, like while playing games, viewing images in a gallery, or reading a book.}. In both approaches, all persistent system bars are hidden. The difference between them is how the user brings the bars back into view, as shown in Figure \ref{fig:screen} (B). 
In \textit{Lean Back} mode, users can bring back the bars by simply touching the screen anywhere. Whereas, in \textit{Immersive} mode, they can just swipe from any edge where a system bar is hidden. This gestures are easy and intuitive, an explanatory message appears on screen the first time the app goes full screen.

\section{Security Model}

When designing \system, we focus on defending against any process running an app that has permissions to access security-sensitive, on-board I/O devices (e.g., camera, microphone, and touch screen) based on the following threat model.

We assume that adversaries have no control over the operating system (e.g., Linux kernel and Android OS) or Android system apps (e.g., SystemUI) and services (e.g., Binder, AudioSystem, MediaServer, SensorManager, InputManager and WindowManager).  Next, we assume that SEAndroid enforces mandatory access control over access to on-board I/O devices, preventing unauthorized access from apps containing native code.  Therefore, we assume that only system services can access on-board I/O devices indirectly through the use of the Java Native Interface (JNI) \cite{JNI} to the Linux kernel.  Further, we assume apps can access I/O devices \textit{only} through APIs provided by the standard Android SDK \cite{AndroidDoc}.  Finally, we assume that Android system services can enforce access control policies that are applied at the time that data would be collected from an I/O device only.  We aim to control whether apps can receive data produced by on-board I/O devices, but do not provide any guarantee after the app is granted access to the data itself.

Adversaries in control of apps may cause threats by executing the following operations.  First, adversarial apps request use of security-sensitive operations on on-board I/O devices, including accessing devices' camera to take picture or video recording, devices' microphones to record users' voices or surrounding environments, and devices' screens to steal security sensitive information displayed to users.  Additionally, adversarial apps may alter the display to launch social-engineering attacks to trick users into consenting to operations they do not want by overlaying a frame buffer over another apps' or system component's display or displaying a request for one operation then performing a different operation.

\section{\system Design Overview}
\label{overview}

\subsection{\system Operation}
\label{operation}

\begin{figure} [t!]
\centering
\includegraphics[width=75mm]{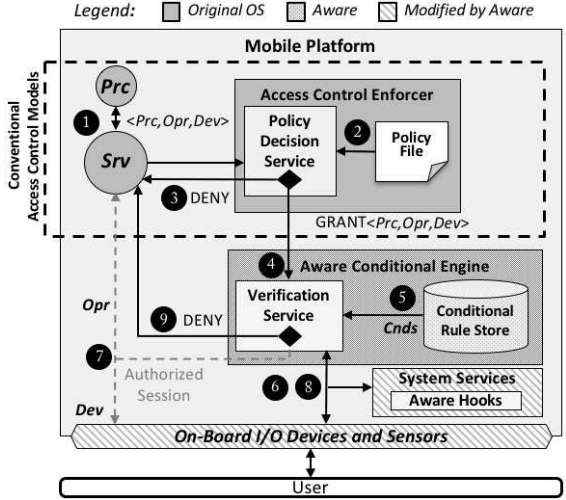}
\caption{Overview of the \system Design}
\label{fig:overview}
\end{figure}

An overview of the \system design is shown in Figure~\ref{fig:overview}. In the overview description, we use Android OS as reference, similar considerations hold for other mobile operating systems available on the market. 
Typically, a process $Prc$, running an Android app, sends a request to perform an operation $Opr$ using a specific I/O device $Dev$ (step~\circled{1}). An example could be a request from an app ($Prc$) to access the front camera ($Dev$) and take a photo ($Opr$). The request is received by a system service $Srv$ (e.g., the MediaServer for the camera), one of the privileged processes in charge of authorizing and processing access requests to system resources. At first, the conventional access control mechanisms are activated (step~\circled{2}). For instance, in Android OS, the Android permissions and the SE Android access control policy checks are activated. If the result of the conventional access control enforcement is a denial, then the system service $Srv$ is notified of the security exception, which in turn notifies the requesting process $Prc$ of the access denial (step~\circled{3}). Otherwise, if the result of the conventional access control enforcement is a grant, \system is activated in order to implement its additional access control conditional checks.

The \system Conditional Engine collects information about the process $Prc$ requesting a certain operation $Opr$ over a target device $Dev$ (step~\circled{4})
%, and authorized to access it, 
through the system service $Srv$.
%, by the conventional access control mechanisms. 
Based on this information, the \system Conditional Engine then identifies and selects, from the Conditional Rule Store, the set of conditional rules $Cnds$ that must be satisfied to allow operation $Opr$ to be performed on behalf of process $Prc$, over the target device $Dev$, by the system service $Srv$ (step~\circled{5}). Subsequently, \system collects the measurements necessary to evaluate if the selected conditional rules are satisfied (step~\circled{6}). Measurements can involve readings from on-board I/O devices and sensors, or system events necessary to verify specific environmental conditions. % (i.e., whether the user is pressing down a soft-button on screen). 
%The set of Conditional Rules are divided into three groups: (1) Pre-Conditions, that must be satisfied before the operation can be authorized; (2) Ongoing-Conditions, that must be satisfied while the operation is performed. 
If and only if \textit{all} the selected conditional rules $Cnds$ are satisfied, then \system generates an authorized session during which the system service $Srv$ is authorized to perform the security-sensitive operation $Opr$, over the target device $Dev$, on behalf of process $Prc$ (step~\circled{7}). During authorized sessions, \system verifies that all conditional rules remain satisfied during the entire session and notifies users about ongoing operations % (i.e., by means of visual hints visible on the device screen) 
(step~\circled{8}). Whenever \textit{any} of the selected conditional rules is not satisfied, \system abort the operation $Opr$ and notifies the service $Srv$ about the set of unsatisfied conditional rules via a denial message (step~\circled{9}). 

\begin{table*}[t!]
\centering
\label{tab:rules}
\setlength{\tabcolsep}{.2em} % REDUCE HORIZONTAL PADDING

\scriptsize
\caption{\system Conditional Rules and Security Properties}
\label{rules}
\begin{tabular}{|l|l|}
\hline
\rowcolor[HTML]{C0C0C0} 
Preconditions        & Security Properties                          \\ \hline
\begin{tabular}[c]{@{}l@{}}\textcolor{white}{\hlc[black]{P1}} User interacts with process \textit{Prc} to request operation \textit{Opr} targeting device \textit{Dev} 
\\ \textcolor{white}{\hlc[black]{P2}} Process \textit{Prc} requests Service \textit{Srv} to perform operation \textit{Opr} over device \textit{Dev}
\\ \textcolor{white}{\hlc[black]{P3}} User is aware of what operation \textit{Opr}, targeting device \textit{Dev}, is going to be\\ \hspace{15pt}performed  by Service \textit{Srv} on behalf of process \textit{Prc} 
\\ \textcolor{white}{\hlc[black]{P4}} User approves operation \textit{Opr} on behalf of process \textit{Prc} targeting device \textit{Dev} 

\end{tabular}   &
\begin{tabular}[c]{@{}l@{}}{\scriptsize{\fcolorbox{gray}{gray}{\bf{SP1}}}} All app requests to perform security-sensitive operations targeting \\\hspace{20pt}  on-board I/O devices must be authorized \\ %Security-sensitive operations targeting on-board I/O devices are performed\\ \hspace{20pt} only when initiated by users \\ 
{\scriptsize{\fcolorbox{gray}{gray}{\bf{SP2}}}} Security-sensitive operations performed using on-board I/O devices \\ \hspace{20pt} match a users' consenting action 

%Security-sensitive  operations  performed  over  on-board  I/O  devices  match\\ \hspace{20pt}  users'  intention  and  volition
\end{tabular}
\\ \hline
\rowcolor[HTML]{C0C0C0} 
Ongoing Conditions                              \\ \hline
\begin{tabular}[c]{@{}l@{}}\textcolor{white}{\hlc[black]{O1}} User has continuous visibility of operation \textit{Opr} performed\\ \hspace{16pt} on behalf of process \textit{Prc} over device \textit{Dev}\\ \textcolor{white}{\hlc[black]{O2}} The authorized session \textit{Ses}, for process \textit{Prc} to execute operation \textit{Opr}\\ \hspace{16pt} over device \textit{Dev}, is logged to allow retroactive actions\end{tabular}        &
\begin{tabular}[c]{@{}l@{}}{\scriptsize{\fcolorbox{gray}{gray}{\bf{SP3}}}} Ongoing security-sensitive operations targeting I/O devices are always  \\ \hspace{20pt} visible  to  users \\  {\scriptsize{\fcolorbox{gray}{gray}{\bf{SP4}}}} Ongoing security-sensitive operations targeting I/O devices are always \\ \hspace{20pt}logged
\end{tabular}           \\ \hline
\rowcolor[HTML]{C0C0C0} 
Exit Conditions                                 \\ \hline
\begin{tabular}[c]{@{}l@{}}\textcolor{white}{\hlc[black]{E1}} The authorized session \textit{Ses}, relative to process \textit{Prc}, terminates \\ \textcolor{white}{\hlc[black]{E2}} Termination of session \textit{Ses} is logged
\\ \textcolor{white}{\hlc[black]{E3}} User has visibility that session \textit{Ses} has been terminated\end{tabular} &
\begin{tabular}[c]{@{}l@{}}  {\scriptsize{\fcolorbox{gray}{gray}{\bf{SP5}}}} All  on-going security-sensitive operations targeting I/O devices \\\hspace{20pt}are visible to the user as long as they run
\end{tabular}                                                                                                                                                    \\ \hline
\end{tabular}  
\end{table*}

\subsection{\system Conditional Engine}

\system Conditional Engine uses \textit{conditional rules} to authorize app requests to I/O devices and maintain security state during any authorized sessions.  To verify the satisfaction of conditional rules, \system is designed  to collect inputs from I/O devices, sensors and system services at run-time, by using additional hooks placed inside the Android framework. 
%Hooks are activated based on the set of conditional rules active in the Conditional Rule Store.
\system conditional rules are of three types: (1) \textit{Preconditions} that must be satisfied before an operation targeting I/O devices is authorized; (2) \textit{Ongoing Conditions} that must be satisfied while the authorized operations targeting I/O devices are being performed, until their completion; and (3) \textit{Exit Conditions} that must be all satisfied once authorized operations terminate due to users' actions or runtime exceptions. 

\system \textit{Preconditions} ensure two security properties. {\scriptsize\fcolorbox{gray}{gray}{\bf{SP1}}} All app requests to perform security-sensitive operations targeting on-board I/O devices must be authorized. This security property prevents processes from performing such operations stealthly. {\scriptsize\fcolorbox{gray}{gray}{\bf{SP2}}} Security-sensitive operations performed using on-board I/O devices match a users' consenting action. This property ensures that every such operation is initiated by a user action.  
%users are aware of the effect of their interaction with a soft-buttons displayed on screen by apps. Third, to ensure that the operation performed by the application in response of the interaction by the user is consistent with the user volition and expectation.
\system \textit{Ongoing Conditions} ensure two additional security properties. {\scriptsize\fcolorbox{gray}{gray}{\bf{SP3}}} Ongoing security-sensitive operations targeting I/O devices are always visible to users.  
This security property enables users to check that the authorized operation is what they expected, reducing the possibility of undetected social-engineering attacks.  %Note that users do not need to track the status of an ongoing operation, but are simply able to check that the expected operation is being run.
{\scriptsize\fcolorbox{gray}{gray}{\bf{SP4}}} Ongoing security-sensitive operations targeting I/O devices are always logged.  Such logging enables users to examine the progress of ongoing security-sensitive operations, which may enable termination of an ongoing operation deemed suspicious or retrospective analysis of past operations.  %reduce the success of social-engineering attacks and identify attempts to access I/O devices from malicious apps, as explained in Section \ref{logs}.
%Condition rules are expressed in form of Boolean Predicates. They evaluate current environmental or system status to check whether relevant requirements are satisfied or not and return either true or false. 
\system \textit{Exit Conditions} ensure one more security property. {\scriptsize\fcolorbox{gray}{gray}{\bf{SP5}}} All ongoing security-sensitive operations targeting I/O devices are visible to the user as long as they run.  This property ensures that apps cannot keep operations running after user terminates them and the user interface correctly removes only terminated operations from the display.
%Table \ref{rules} reports \textit{Preconditions}, \textit{Ongoing Conditions} and \textit{Exit Conditions} adopted in our current system prototype to ensure the satisfaction of the five security properties mentioned above. 
The conditional rules and security properties above mentioned are summarized, for future reference, in Table \ref{rules}. To express \system conditional rules, the Usage Control (UCON) model \cite{park2004ucon} could be adopted.

\vspace{-3mm}

\section{\system Design}

We present the \system design in terms of four mechanisms necessary to fulfill the five security properties.  

\subsection{Mediation of Access Requests}
\label{mediate}

%\subsection{Environment and System Event Monitoring}

%problem in achieving goals
Complete mediation of \textit{all} requests to access I/O devices from processes running apps and matching each request to a user input corresponding to the app request is necessary to ensure that only authorized app requests are run, guaranteeing {\scriptsize\fcolorbox{gray}{gray}{\bf{SP1}}}. %Furthermore, internal system events must be combined to identify users' volition when interacting with apps. 
Mediation involves several system services, such as services controlling gestures on screen, or handling requests from running processes, when targeting on-board I/O devices and sensors. Users' interaction events have to be aggregated and mapped to requests instantiated by processes running apps. The SE Android reference monitor must be extended to ensure {\em complete mediation} of all security-sensitive operations targeting I/O devices~\cite{monitor}.  Identifying the right locations where to place additional hooks, to mediate every access to I/O devices, is challenging.

%design alternatives
Mainly, complete mediation of accesses to I/O devices can be achieved in two ways: (1) by placing hooks inside the Android framework and libraries, or (2) by placing hooks inside the Linux kernel and I/O device drivers. To achieve complete mediation of accesses to I/O devices, which are low-level system resources, the kernel and device drivers seem to be the most appropriate place where to add mediation hooks. However, two main issues arise from this approach: (1) low level hooks would not have the required level of information to map requests to processes running apps, due to the fact that requests are always handled by system services on behalf of the requesting processes; (2) mobile platforms are equipped with different I/O devices, which would require the operating system to be able to support customized hooks defined for different drivers by driver vendors.

%prove design choice are superior to others
In \system, hooks are placed at the Android framework and libraries level, to avoid the above mentioned issues. \system \textit{Hooks} provide complete mediation, because system services are the only path through which processes, running apps, can access I/O devices and sensors, due to Android framework architecture and MAC rules enforced by SE Android \cite{AndroidEnh}. We have dynamically analyzed the Android framework and libraries code, relative to SDK APIs handling accesses to I/O devices and sensors, to validate complete mediation, and check that every access to the I/O devices and sensors is captured by one of the 18 hooks introduced. Retaining such logging could be used to detect errors, if any exist. Callbacks from hooks inform the \system Conditional Engine about users interacting with processes running apps, and requesting operations over I/O devices. These callbacks are used to validate precondition{\small\textcolor{white}{\hlc[black]{P1}}}. \system \textit{Hooks} also capture resources acquisition and release by system services operating on behalf of processes running apps. Callbacks from these hooks are used to validate precondition {\small\textcolor{white}{\hlc[black]{P2}}} and exit condition {\small\textcolor{white}{\hlc[black]{E1}}}. Satisfying these conditions is sufficient to reliably bind users interaction with apps requests to operate over I/O devices, therefore guaranteeing {\scriptsize\fcolorbox{gray}{gray}{\bf{SP1}}}.

\subsection{Visibility over Sensitive Operations} \label{secure_message}

%\subsection{Conveying Access Information to Users} \label{secure_message}

%problem in achieving goals
Secure display of operations targeting I/O devices when they are requested, when they are ongoing, and when they are terminated is necessary to fulfill multiple guarantees.  First, by maintaining visibility of operations after they are authorized, users may identify undesired operations approved by mistake to guarantee {\scriptsize\fcolorbox{gray}{gray}{\bf{SP3}}}. Furthermore, ensuring that operations are visible to users as long as they run guarantees that there are no stealthy operation on I/O devices ongoing  {\scriptsize\fcolorbox{gray}{gray}{\bf{SP5}}}.
Visibility over accesses to I/O devices from running apps may be provided to users in four different ways: (1) via notification lights, similar to those used for cameras on laptops or external USB cameras; (2) by playing a distinctive sound, similar to the shutter sound produced when taking a photo; (3) by displaying notification icons, similar to the location icon shown on the status bar; and (4) by visualizing alert messages on screen. Unfortunately, notification lights, sounds, or notification icons can only alert users about accesses to sensitive I/O devices, but cannot convey exact information about operations performed, target devices, and processes responsible for such operations. Furthermore, the sounds might not be audible in silent or vibrate mode. A better way to convey complete information about operations performed over I/O devices, by running processes, is by displaying on screen alert messages to users.

Solutions that make use of the \textit{Activity Window} portion of the screen, to display access notifications or alert messages, are subject to user deception attacks, were screen overlays are used by malicious apps to surreptitiously replace, or mimic, the GUI of other apps and mount social-engineering attacks, or else mislead the user perception of ongoing operations. 
% based on the well-known HTTPS lock icon used in current browsers

\system avoids this problem by displays \textit{Security Messages} to users on the \textit{Status Bar}, a reserved portion of the screen drawable only by the \textit{WindowManager} system service, a privileged process part of the Android OS.  A similar approach has been adopted by Bianchi \textit{et al.} \cite{bianchi2015app}, where the \textit{Navigation Bar} has been used to host a security indicator as solution against User Interface (UI) deception. However, Bianchi's solution, as it is, cannot be adopted to provide visibility to users about operations targeting I/O devices or to automatically prevent operation programmatically initiated by processes running apps, without users interaction. In fact, Bianchi's solution does not provide the necessary mechanisms to bind users interaction to access requests for I/O devices from processes running apps.

\begin{figure}[]
\centering
\includegraphics[width=60mm]{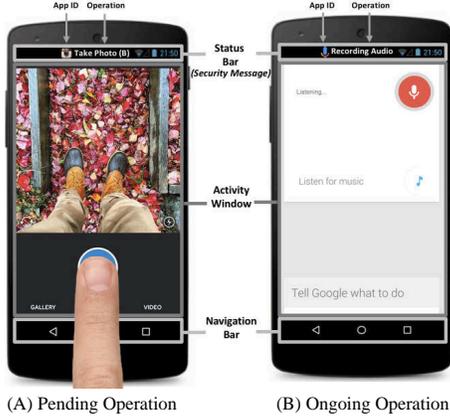}
\\\scriptsize{(A) Pending Operation \hspace{12mm} (B) Ongoing Operation}
\vspace{-3mm}

\caption{Security Messages displayed on the Status Bar}

\label{fig:messages}
\end{figure}

\system uses \textit{Security Messages} displayed on the \textit{Status Bar} to convey, to users, two types of events:
(1) \textit{pending operations} initiated by users; and (2) status feedback about \textit{ongoing operations} authorized by users. A \textit{Security Message} includes the app identifier (e.g., app icon  or name) and a text message specifying the target operation. 
The first type of message makes users aware of the  operation resulting from the interaction with a soft-button displayed by an app on screen.
For example, in Figure \ref{fig:messages} (A), if the user presses the button depicting a camera, the \textit{Security Message} specifies that the Instagram app will take a photo using the smartphone front camera\footnote{(F) used to indicate front camera and (B) indicate back camera.}.  The second type of message informs users about ongoing authorized sessions targeting on-board I/O devices. As example, in Figure \ref{fig:messages} (B), a \textit{Security Message} is used to inform the user that the Google Voice Search app is using the microphone to listen to the user voice for commands. 
If multiple operations are simultaneously targeting different I/O devices, the \textit{Security Message} alternate messages to make users aware of all the ongoing operations.

\textit{Security Messages} are used to validate precondition{\small\textcolor{white}{\hlc[black]{P3}}}, ongoing condition {\small\textcolor{white}{\hlc[black]{O1}}}and exit condition{\small\textcolor{white}{\hlc[black]{E3}}}. Satisfying these conditions is sufficient to reliably provide visibility to users over sensitive operations targeting I/O devices, therefore guaranteeing  {\scriptsize\fcolorbox{gray}{gray}{\bf{SP3}}} and
{\scriptsize\fcolorbox{gray}{gray}{\bf{SP5}}}.

\system \textit{Security Messages} are always visible to users %when the foreground app is not in full screen mode. On the other hand, 
even if apps are in full screen mode. Upon use-initiated operations targeting I/O devices (i.e., press soft-button to take a photo), \system systematically reactivates the \textit{Status Bar} to display a \textit{Security Message} specifying the pending operation. Thus, any attempt by malicious apps to draw a fake \textit{Status Bar} with a fake \textit{Security Message} would fail, since the original \textit{Status Bar} is always drawn on screen.

\subsection{Eliciting User Input for Approval} \label{user_input}

The requirement for users to approve or abort pending app requests for operations on I/O devices by providing user input through  GUI elements, guarantees {\scriptsize\fcolorbox{gray}{gray}{\bf{SP2}}}.
%Any operation targeting I/O devices must be initiated by the user by interacting with the app requesting the operation itself, therefore, a per-access approval mechanism must be used to ensure that operations requested by running processes correspond to users' volition.
%design alternatives
On-screen prompts could be used to request approval, every time I/O devices are accessed by a process running an app, in response to  user-initiated interactions.  However, while prompts attempt to verify users' intention, in practice, they create an excessive burden on users, which leads to users ignoring these prompts eventually\footnote{On average, there are 8 requests per minute by processes running apps to request permission to access sensitive resources \cite{primal}.}. Therefore, prompting users every time I/O devices are accessed seems unreasonable.

\begin{figure}[t!]
\centering
\includegraphics[width=80mm]{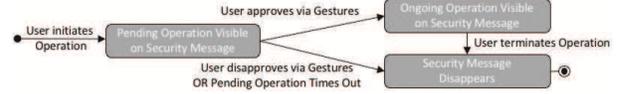}
%\scriptsize{(A) \hspace{17mm} (B) \hspace{22mm} (C) \hspace{7mm}}
%\vspace{-3mm}
\caption{State Transition Diagram for Security Messages based on User-Initiated Interactions}
\label{fig:stm}
\end{figure}

\begin{figure}[t!]
\centering
\includegraphics[width=80mm]{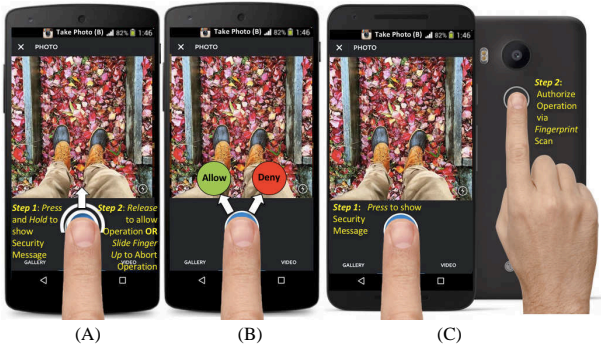}
\scriptsize{(A) \hspace{17mm} (B) \hspace{22mm} (C) \hspace{7mm}}
\vspace{-3mm}
\caption{Alternative Approaches to retrieve Users Input}
\label{fig:user_input}
\end{figure}

%prove design choice are superior to others
To avoid excessive burden on users and, at the same time, enforce a per-access approval, \system uses a \textit{Gesture Identification} mechanism to identifying specific sequence of gestures, by users, on smartphones' screen. Gestures are intercepted and analyzed, in real-time, to infer users' intention when interacting with apps. Captured gestures on screen are mapped with undergoing operations performed by apps running in foreground. User-initiated interactions are combined with \textit{Security Message} on screen, as depicted in the state machine diagram in Figure \ref{fig:stm}. With \system, operations targeting I/O devices can \textit{only} be initiated by user, pressing and holding down a soft-button on screen. 
Upon user-initiated operations, \system displays the pending operation on a \textit{Security Message}, for a preset time period, after which the operation is abort in absence of user interaction\footnote{The timer is used to support apps that require users to keep pressing down a button to perform the operation (i.e., record a video).}. After looking at the  \textit{Security Message} users can confirm the operation by simply releasing the soft-button, or aborting the pending operation by sliding their finger out from the soft-button area, as shown in Figure \ref{fig:user_input} (A).

We could have designed \system in order to have two different areas of the screen where users could place their finger to either deny or allow operations, as shown in figure \ref{fig:user_input} (B). This solution would have been subject to social-engineering attacks because the two areas would appear in the \textit{Activity Window}, allowing malicious apps to overlay fake messages, and swap the deny area with the allow area to trick the user into allowing an operation.

The \system \textit{Gesture Identification} mechanism also support an alternative method that makes use of the fingerprint scanner to authenticate users interacting with smartphones%\footnote{Appealing alternative since new smartphones (e.g., iPhone 5s and Nexus 5X) are being equipped with a fingerprint scanner, originally meant to support payment and lock screen methods.}. 
Users scan their finger to confirm pending operations displayed on \textit{Security Messages}, as illustrated on the right side of Figure \ref{fig:user_input} (C). \system interprets the absence of specific sequences of gestures, from users, how operations not matching users' intention and volition, therefore, blocks and logs attempts from malicious apps trying to programmatically activate security-sensitive operations targeting on-board I/O device. The \textit{Gesture Identification} mechanism is used to validate precondition {\small\textcolor{white}{\hlc[black]{P4}}}, which in conjunction with  {\small\textcolor{white}{\hlc[black]{P1}}}, {\small\textcolor{white}{\hlc[black]{P2}}} and {\small\textcolor{white}{\hlc[black]{P3}}} are sufficient to guarantee that operations performed over on-board I/O devices match users' intention and volition,  therefore guaranteeing {\scriptsize\fcolorbox{gray}{gray}{\bf{SP2}}}.

For users willing to lower the security of their mobile platforms, \system allows to disable the \textit{Gesture Identification} mechanism per-app or when a remote controller (i.e., Bluetooth selfie stick) is used. However, we discourage white listing apps, even after a certain period of usage, because apps can dynamically change their behavior during time, due to automatic, periodic, software updates. Furthermore, apps could ask another apps to perform specific operations targeting I/O devices, through the intent mechanism. Thus, a white listed app could be tricked in serving a request coming from a malicious app.

\subsection{Supporting Retrospective Actions} \label{logs}

\begin{figure}[h]
  \centering\includegraphics[width=75mm]{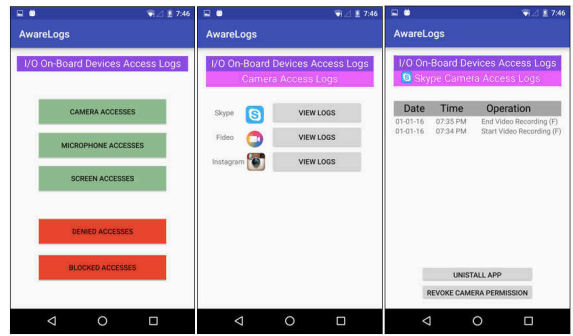}
  \caption{On-Board I/O Devices Access Logs}
\label{fig:logs}
\end{figure}

 The requirement to log actions occurring during the execution of operations  targeting I/O devices, guarantees  {\scriptsize\fcolorbox{gray}{gray}{\bf{SP4}}}. To support retroactive actions by users,  \system generates three type of access logs for security-sensitive operation targeting on-board I/O devices. First, \system logs any failed attempt by running processes in accessing I/O devices, due to lack of necessary conditions required to allow requested operations. These logs are accessible in the \textit{Blocked Accesses} section, shown in Figure \ref{fig:logs}, and allow users to identify apps that attempts to perform stealthy operations while running as background services. Second, \system logs any operation denied by users though the \textit{Gesture Identification} mechanism. These logs are accessible in the \textit{Denied Accesses} section, shown in Figure \ref{fig:logs}, and allow users to identify apps using social-engineering techniques to trick them in authorizing undesired operations. Third, \system log any operation performed over I/O devices, authorized by users though the \textit{Gesture Identification} mechanism, allowing users to track authorized operations. 
 
 To better catch users' attention, attempted access violations are signaled by \system by producing a sound and showing a \textit{Security Message} communicating undesired behaviors from running apps. The \system \textit{Logs} can be accessed by users anytime, from the app menu or by tapping on the \textit{Security Message} displayed on the \textit{Status Bar}. Each access log entry reports information regarding apps ID, date, time and operations performed by apps, as shown in Figure \ref{fig:logs}.

\system \textit{Logs} allow users to perform two retrospective actions: (1) uninstall apps identified as malicious, and (2) revoke granted permissions to prevent future undesired accesses\footnote{This mechanism supports the Android AppOps mechanism reintroduced starting from Android OS 6.0 Marshmallow \cite{AndroidDoc}, which allows to revoke permissions granted at install time, to running apps.}. Retrospective actions can be taken by users either immediately, when the I/O devices is still being used by apps to block undesired operations, or after reviewing what apps are doing over time.

The \system \textit{Log} mechanism is used to validate Ongoing Condition{\small\textcolor{white}{\hlc[black]{O2}}} and Exit Condition{\small\textcolor{white}{\hlc[black]{E2}}}. Satisfying these conditions is sufficient to allow users to perform retrospective actions over sensitive operations targeting I/O devices, therefore guaranteeing {\scriptsize\fcolorbox{gray}{gray}{\bf{SP4}}}.

\input{implementation}

\section{User Study and Aware Evaluation}

\subsection{Study Objectives}

We performed a comprehensive laboratory-based survey, user study, and system experiments with the following five objectives. First, we survey users' privacy and security attitudes as they pertain to the malicious use of I/O devices, and investigate users' awareness about RAT attacks. Second, we observe users' vigilance during a series of interactive tasks, while RAT attacks targeting on-board I/O devices are deployed and the \system defense mechanisms are not active. Third, we investigate the effectiveness of \system from two perspectives. Initially, during another series of interactive tasks, we observe whether users notice and can adequately respond to customized social-engineering attacks, when those can be thwarted with the user interface components of \system. We then debrief users about their experience with \system. Fourth, we measure whether \system effectively and systematically shields users from practically deployed RAT attacks, while they perform the second series of interaction tasks.  Fifth, we measure the performance overhead \system incurs on the critical paths of processing app requests.

\begin{table*}
\centering
\label{task}
\setlength{\tabcolsep}{.5em} % REDUCE HORIZONTAL PADDING

\scriptsize
\caption{List of Experimental Tasks and Attacks Description}
\label{task}

\begin{tabular}{|c|l|l|l|l|l|c|}
\hline
{\color[HTML]{333333} \begin{tabular}[c]{@{}l@{}}\textbf{Task ID}\end{tabular}} & {\color[HTML]{333333} \begin{tabular}[c]{@{}l@{}}\textbf{Task Description}\end{tabular}} & {\color[HTML]{333333} \begin{tabular}[c]{@{}l@{}}\textbf{App Used}\end{tabular}} & {\color[HTML]{333333} \begin{tabular}[c]{@{}l@{}}\textbf{Attack Source}\end{tabular}} & {\color[HTML]{333333} \begin{tabular}[c]{@{}l@{}}\textbf{Attack Description}\end{tabular}} & {\color[HTML]{333333} \begin{tabular}[c]{@{}l@{}}\textbf{Perceivable}\end{tabular}} & {\color[HTML]{333333} \begin{tabular}[c]{@{}l@{}}\textbf{Detection Rate}\end{tabular}} \\ \hline 
\cellcolor[gray]{0.9}T1 & Take a picture  & Instagram (B) & None  & N/A & N/A  & \cellcolor[gray]{0.95}N/A  \\\hline
\cellcolor[gray]{0.9}T2 & Take a video  & Fideo (B) & None  & N/A & N/A  & \cellcolor[gray]{0.95}N/A \\ \hline
\cellcolor[gray]{0.9}T3 & Record a voice message  & Messenger (B)  & None & N/A & N/A & \cellcolor[gray]{0.95}N/A \\ \hline
\cellcolor[gray]{0.9}T4 & Record a video message & Skype (B) & None & N/A & N/A  & \cellcolor[gray]{0.95}N/A \\ \hline
\cellcolor[gray]{0.9}T5 & Record the device screen  & Rec. (B) & None  & N/A & N/A  & \cellcolor[gray]{0.95}N/A \\ \hline
\cellcolor[gray]{0.9}T6-A1 & Navigate Internet & Chrome (B) & Krysanec  (M) & Stealthy Photo & Camera Shutter Sound & \cellcolor[gray]{0.95} 18\% \\ \hline
\cellcolor[gray]{0.9}T7-A2 & Watch a Video & YouTube (B) & Soundcomber (M) & Stealthy Voice Recording & None & \cellcolor[gray]{0.95}0\%\\ \hline
\cellcolor[gray]{0.9}T8-A3 & Add a new contact  & Contacts (B)  & Dendroid (M)  & Stealthy Video Recording & UI Slow Down & \cellcolor[gray]{0.95}1\% \\ \hline
\cellcolor[gray]{0.9}T9-A4 & Send email  & Gmail (B) & PlaceRaider (M) & Stealthy Photos & UI Slow Down & \cellcolor[gray]{0.95}0\%\\ \hline
\cellcolor[gray]{0.9}T10 & Take a screenshot & None* & None & N/A & N/A & \cellcolor[gray]{0.95}N/A\\ \hline
\cellcolor[gray]{0.9}T11-A5 & Record a video & SimpleFilters $\dagger$ &  SimpleFilters $\dagger$ & Stealthy Screenshot & Security Message Mismatch & \cellcolor[gray]{0.95}82\%\\ \hline
\cellcolor[gray]{0.9}T12-A6 & Take a photo  & SimpleFilters $\dagger$  & SimpleFilters $\dagger$   & Stealthy Voice Record & Security Message Mismatch & \cellcolor[gray]{0.95}76\% \\ \hline
\cellcolor[gray]{0.9}T13 & Analyze \system Logs  & \system Logs  (B) & None  & N/A & N/A & \cellcolor[gray]{0.95}N/A\\ \hline
\end{tabular}
\\
{\scriptsize (B) Benign App \hspace{5mm} (M) Malicious App \hspace{5mm}* Performed by pressing the power and volume-down physical buttons at the same time 
\\ $\dagger$ SimpleFilters appears as a benign app, but includes functionality to run additional I/O operations beyond those consented by users}
%, until a click or screenshot sound is produced.
\end{table*}

\subsection{Study Components}

The study has three survey components, two interactive user task sequences, and one group of measured tasks. We obtained IRB approval at our institution. 

\textit{\underline{Surveys}:} Individuals completed an initial questionnaire with demographic questions and questions about their usage of mobile platforms. A second survey debriefed participants about the first series of interactive tasks, performed using an of-the-shelf Android smartphone and investigated their privacy and security attitudes. A third survey debriefed participants about the second series of interaction tasks, performed using a smartphone running \system, and their perceptions about \system. The surveys included standard Likert-type psychometric scale questions (e.g., to measure attitudes) as well as open-ended question formats (e.g., to solicit participants' experiences during the interactive tasks). Surveys were implemented on Qualtrics and executed on a lab computer.

%, given the security features of the vanilla Android OS

\textit{\underline{Interactive Tasks}:} 
In the first series of interactive tasks, we studied participants' potential reactions to practically deployed RAT attacks. Participants were asked to interact with a Nexus 5 smartphone, running a vanilla version of the Android OS (6.0.1\_r25), and to perform 9 tasks ranging from taking a picture with the smartphone's camera to sending an email. The first 5 tasks (T1-T5), summarized in Table \ref{task}, were not associated with any RAT attacks. Tasks T6-T10 were associated with 4 different, visibly noticeable attacks (A1-A4), also summarized in Table~\ref{task}. These attacks were carefully triggered by the experimenter, while participants engaged in the interactive tasks. The attacks varied in the degree to which they are perceivable, as highlighted in Table~\ref{task}. Please note that individuals were not explicitly instructed about the presence of RAT attacks before the tasks, however we asked them to report unusual behaviors verbally and in the survey. %As such, this part aims to understand the effectiveness of RAT attacks targeting on-board I/O devices, when the smartphone is running an off-the-shelf mobile operating system. %We summarize the 9 tasks and associated attacks in Table~\ref{task}. 

Before the second series of interactive tasks, subjects were debriefed about the previously experienced attacks. We further familiarized them with the \system system through instructional materials and by allowing them to inspect the \system user interface on a Nexus 5X smartphone. 
%running a modified version of the Android OS (6.0.1\_r25) with the \system system. 
As before, the participants were engaged in several interactive tasks. First, participants engaged in tasks T1-T5 and T10, which did not present any \textit{noticeable} RAT behaviors. Then, in tasks T11 and T12, we used a test RAT app to investigate users' responses to attacks where the user consented to one action, but the app performed additional, unapproved actions. These social-engineering attacks (A5 and A6) aimed to trick users into executing unwanted I/O operations, such as recording voice when they consented to the app taking a photo. Using the \system security message and gesture mechanism, we investigated whether participants could notice and thwart the attacks. The series of tasks concluded with T13, which did not include any attacks. We did not brief individuals about which tasks were associated with attacks or which I/O devices would be targeted. We summarize the 9 tasks and associated attacks in Table~\ref{task}. %This second series of tasks is meant to measure the effectiveness of the \system system in preventing the user in being tricked by malicious app and being victim of a social-engineering attack (i.e., the app shows a camera button on screen but instead of taking a photo it actually record audio).

We recorded participants' interactive behaviors, and their survey responses. In addition, we applied a think-aloud protocol by encouraging participants to speak out about their experiences while being engaged in the tasks.

\textit{\underline{Measured Evaluation}:} During the second series of tasks, participants were still exposed to attacks programmatically and persistently triggered from the four RAT apps used in our study (i.e., Krysanec, Soundcomber, Dendroid, and PlaceRaider). The \system system was expected to shield the user from these attacks, and therefore no \textit{noticeable} effects should have been observable by the participants. As such, the purpose of the measurement task is to evaluate whether \system effectively and automatically shields the participants \textit{while they engage in realistic user interaction behaviors}. 
% Refer to Biachi's Section VII if needed

All participants completed the entire set of study components in about 25-35 minutes and were compensated with a \$10 gift card.

\subsection{Results}

\hspace{2mm} \textit{ \underline{Demographics and Mobile Platforms Usage}:} In total, 74 participants completed the whole set of surveys and tasks. The majority of the sample were between 20-29 years old (76\%). We recruited predominantly undergraduate and graduate students; the majority having an international background (70\%), and fields of study \textit{different} from computer science (75\%). Most participants actively used smartphones (99\%) and additional devices associated with third-party apps such as tablets (54\%), and to a lesser extent smart watches and fitness bands (10\%). 

\textit{ \underline{Privacy Attitudes}:} We asked participants how concerned they are about threats to their personal privacy when using smartphones, and found that 43\% were moderately or extremely concerned. Participants were even more concerned about privacy and security aspects as they related to third-party apps (57\%). Most important to our study, concern levels were high for the misuse of smartphones' camera (62\%) and microphone (55\%).

\textit{ \underline{RAT Awareness and Security Behaviors}:} The majority of the participants stated that they were aware that apps could access the camera (56\%) and the microphone (56\%) of their smartphones at any time without repeatedly asking for consent. However, participants had little knowledge of specific RATs that exploit smartphones' I/O devices, such as Dendroid and SoundComber (4\% each), and Krysanec and PlaceRaider (3\% each). A small number of participants (8\%) were able to articulate how malicious apps apply social-engineering techniques to misled users into taking an action. % (e.g., draw on top of a legitimate app).
Only 24\% of the participants use a mobile anti-malware product, whereas 78\% of subjects stated that they avoid downloading apps from unofficial app stores. 

\textit{ \underline{Identification of Threats without \system}:} The attacks in the first series of interaction tasks varied in the degree to which \textit{risk signals} in the vanilla Android system are perceivable by a user. A1 was associated with the camera shutter sound when the Krysanec malware took a stealthy photo, while participants were browsing the Internet. Only 18\% correctly noticed that a camera shutter sound was audible. 8\% incorrectly thought that a screenshot was taken. 4\% merely noticed a sound. Not a single participant stated any suspicion in the survey or the think-aloud comments that malware or a security problem could be responsible for the sound. Two participants questioned whether ``something'' or ``someone was taking a screenshot''. Only one participant noticed the UI slow-down (i.e., a perceived glitch in the UI) in task T8-A3 when a stealthy voice recording was initiated by the SoundComber app. None of the participants identified any suspicious activity when being involved in attack tasks T7-A2 and T9-A4. An overview of the detection rates for the RAT attacks is available in Table \ref{task}.

In summary, audible risk signals (associated with the built-in camera in A1) are noticed by at least some participants, but their interpretation of the risk, ability to identify the cause or at least to question the cause of the unexpected behavior is highly limited. State-of-the-art malware apps engaging in stealthy voice and video recordings remained almost entirely unnoticed. PlaceRaider (A4) which also takes stealthy pictures and sidesteps audible risk signals was not noticed by any participant. 

\textit{ \underline{Interaction of Users with \system}:} Using \system substantially simplifies the problem space of unwanted I/O access in two ways. First, any access to an I/O device which is not associated with a user interaction will be automatically blocked. Thereby, attacks that participants were encountering in the first series of interactive tasks would be automatically thwarted. We will verify this aspect of \system's effectiveness in the subsequent measurement results subsection. Second, only user-intended interactions which explicitly trigger an I/O device will be allowed to proceed. Note, however, that malicious apps could nevertheless try to hijack these user-intended activities. \system also aims to thwart attacks that coincide with such user-intended I/O interactions. For this purpose, \system provides user interface features (i.e., \system security messages and gesture identification) that help to \textit{notice} and to \textit{block} such attacks. In addition, users can \textit{inspect the \system log} to verify which wanted or unwanted I/O operations have taken place.

To investigate the effectiveness of \system when users encounter such sophisticated attacks (which exploit the \textit{context} of users' interactions), we have customized a simple photographic filter app  called SimpleFilters. The app is modified to attempt to take a screenshot (A5) and a voice recording (A6) stealthily when users intentionally engage in a different type of I/O interaction. Using \system, 82\% of the participants noticed the attack while performing task T11-A5, and 76\% noticed the attack while performing task T12-A6. In contrast to the first series of user interactions, these participants were also able to clearly articulate what attacks took place. A typical debriefing message for attack A5 was: ``It was taking a screenshot, while I asked it to take a video'', whereas for attack A6, messages were a variations of: ``\system message said the microphone was being accessed''.

Inspecting the system logs, we were also able to determine how often participants used the \system gestures to \textit{block} the attacks that they noticed. For A5, all of the 82\% of the participants who noticed the attack successfully used the gestures to abort the task. Similarly, all participants who noticed attack A6 succeeded in blocking the attempt to record audio instead of taking a picture. In the final task (T13), we asked individuals to inspect \system Logs to evaluate which I/O access operations had taken place during the second series of interactions. 88\% of all participants found \system Logs helpful in identifying suspicious activities from running apps, and they were clearly able to articulate what attacks had taken place.

After the interaction tasks, we solicited further feedback from the participants. 90\% of the participants found \system more secure than the vanilla Android OS, and 80\% found it as (or more) usable compared to the vanilla Android OS. These are encouraging results since additional security mechanisms often meet with user resistance, for example, because they may distract from the user's primary task. Further, 57\% of the participants said they would prefer the \system notice and gesture mechanism compared to other notification options. For example, only 21\% of the participants preferred to be prompted with a permission dialog at every access. Further, only 10\% of them stated that they would prefer to be asked for permission at install time, and 8\% of them at first use. Most importantly, 99\% of the participants would like \system integrated in their current mobile OS.

\textit{ \underline{Measurements to Evaluate \system}:} %As discussed in the previous results subsection, users benefit from \system in multiple ways. In particular, malicious app activities that are not carefully crafted to exploit intended user interactions with I/O devices will be automatically blocked. This ability of \system entirely removes the need for users to actively handle most state-of-the-art RAT attacks.
To evaluate the effectiveness of \system in the presence of realistic user interactions, we allowed the 4 RAT apps used in the first series of user interaction tasks to also be active during the second series of interaction tasks. We also tested whether activities from the customized SimpleFilters app would be blocked if they did not coincide with the users' interactions with the I/O devices (i.e., in T11 and T12).
In order to monitor whether any of the malicious activities of these RAT apps were successful, we used logcat \cite{AndroidDoc}, the Android logging system, which provides a mechanism for collecting system debug output about activities from various apps and system services. For the 74 sessions involving participants, we found that RAT apps attempted 1080 times to perform stealthy operations targeting I/O devices, but they \textit{never} succeeded in accessing the on-board camera, microphone or screen content, as result of systematically validating preconditions \textcolor{white}{\hlc[black]{P1}} and \textcolor{white}{\hlc[black]{P4}}. In other words, the absence of users' interaction and consent prevented RAT apps from succeeding in performing stealthy operations while running services in the background. Furthermore, based on the logs, there were no run-time exceptions, triggered by the \system components, which could have caused any of the 9 well-known legitimate apps to crash or unexpectedly terminate.

\textit{ \underline{Summary}\footnote{See Appendix \ref{res} for a summary of selected results.}:} \system prevented all attempts from RAT apps to perform stealthy operations that did not coincide with users' intended I/O access operations and considerably reduced the success rate of social-engineering attacks, without breaking any apps' logic. Therefore, \system significantly raises the bar compared to the detection rate of state-of-the-art static/dynamic analysis tools, and anti-malware tools, available to users to identify malicious apps running on smartphones\footnote{Detailed analysis results reported in Appendices \ref{anti} and \ref{stealthystalker}.
}. We anticipate that with additional experience users will become even more proficient with the \system security messages and the gesture mechanism, which would further reduce the effectiveness of social-engineering techniques. However, the achieved results are very impressive given the sophisticated nature of the attacks tested with the SimpleFilters app\footnote{Research on carefully crafted Phishing attacks shows that, even with repeated security training, a significant share of users will fall for such attacks \cite{Sheng10}.}. \system automatically blocks all attacks which are not carefully socially-engineered and significantly reduces the attention burden placed on users, thereby, reduces habituation and notice fatigue \cite{Bohme11}.

\begin{table}[t]
\scriptsize
\centering
\setlength{\tabcolsep}{.1em} % REDUCE HORIZONTAL PADDING

\caption{\system Performance Overhead in $\mu$s. Numbers give the mean value and corresponding standard deviation after 10,000 runs. }
\label{performance}
\begin{tabular}{c|c|c|c|c|
>{\columncolor[HTML]{EFEFEF}}c |}
\cline{2-6}
                                                            & \multicolumn{2}{c|}{\cellcolor[HTML]{9B9B9B}{ Vanilla Android OS}}                                  & \multicolumn{3}{c|}{\cellcolor[HTML]{9B9B9B}{ \system}}                                                                                               \\ \cline{2-6} 
\cellcolor[HTML]{FFFFFF}                                    & \multicolumn{1}{c|}{\cellcolor[HTML]{C0C0C0}Nexus 5} & \multicolumn{1}{c|}{\cellcolor[HTML]{C0C0C0}Nexus 5X} & \multicolumn{1}{c|}{\cellcolor[HTML]{C0C0C0}Nexus 5} & \multicolumn{1}{c|}{\cellcolor[HTML]{C0C0C0}Nexus 5X} & \multicolumn{1}{c|}{\cellcolor[HTML]{C0C0C0}\begin{tabular}[c]{@{}c@{}}Max (Min)\\ Overhead\end{tabular}} \\ \hline
%\multicolumn{6}{|c|}{{\cellcolor[gray]{0.6}On-Board I/O Devices}} \\ \hline 

\multicolumn{1}{|l|}{\cellcolor[gray]{0.9}Front Camera} &  15.90$\pm$1.54 &  14.39$\pm$1.12 & 16.11$\pm$1.77  & 15.01$\pm$1.38 &  4.04\% (2.21\%)  \\ \hline
\multicolumn{1}{|l|}{\cellcolor[gray]{0.9}Back Camera} &  16.08$\pm$1.32 &  15.68$\pm$1.87 & 16.44$\pm$1.06  & 16.37$\pm$1.91 &  4.31\% (2.57\%)     \\ \hline
\multicolumn{1}{|l|}{\cellcolor[gray]{0.9}Microphone} & 12.36$\pm$2.01 &  11.86$\pm$1.99 & 12.65$\pm$2.15  & 12.32$\pm$1.85 &  4.03\% (2.19\%)    \\ \hline
\multicolumn{1}{|l|}{\cellcolor[gray]{0.9}Screen}  & 17.76$\pm$0.99 &  16.23$\pm$0.69 & 18.61$\pm$0.90  & 17.02$\pm$1.01 &  4.79\% (2.94\%)   \\ \hline

\end{tabular}
\end{table}

\subsection{Performance Evaluation}

We have measured the overhead introduced by \system while handling each access request for operating on-board I/O devices, such as the camera to take photos and video, the microphone to record audio, and the screen to capture screenshots. Due to lack of publicly available benchmarks for Android OS, we only provide microbenchmark analysis of such operations for two phones, a Nexus 5 and a Nexus 5X running Android OS (version android-6.0.1\_r5). 
%We also have measured the overhead introduced by \system while handling each access request for operating on-board sensors, such as the gyroscope, the accelerometer and the GPS receiver.
The overhead is calculated  by measuring the time interval from the time the request is made by the process running the app to the time the operation is granted/denied by \system. Table \ref{performance} reports the average time over 10,000 requests, the standard deviation and the maximum recorded overhead introduced by \system. Overall, \system introduces a negligible overhead of the order of 1 $\mu$s per access. The maximum recorded overhead is 4.79\% while accessing the screen buffers. 

\input{related_work}

\section{Conclusion}
 
In this paper, we presented \system, a security framework for authorizing app requests to perform sensitive operations using I/O devices, which binds app requests with user intentions to  make  all  uses  of  certain  I/O  devices explicit. We evaluated the proposed defense mechanisms through laboratory-based experimentation and a user study, involving 74 human subjects, whose ability to identify undesired operations targeting I/O devices increased significantly. Without \system, only 18\% of the participants were able to identify attacks from tested RAT apps. \system systematically blocked all the attacks, in absence of user-initiated interaction, and supported users in identifying 82\% of more sophisticated attacks, which used social-engineering techniques to hijack user-initiated operations. \system introduced only 4.79\% maximum performance overhead over operations targeting I/O devices.

%{\footnotesize \bibliographystyle{acm}
%\bibliography{../common/bibliography}}

%{\footnotesize \bibliographystyle{IEEEtran}
%\bibliography{Ref}}
%\bibliographystyle{plain}

%\addbibresource{Ref.bib}

%appendix
\section*{Appendices}
\addcontentsline{toc}{section}{Appendices}
\renewcommand{\thesubsection}{\Alph{subsection}}

\subsection{Android Permission Set Analysis}

\label{perm_ineff}

\begin{table*}[t]
\scriptsize
\centering
\setlength{\tabcolsep}{.5em} % REDUCE HORIZONTAL PADDING

\caption{Market Apps that potentially could behave as RATs and perform stealthy operations}
\label{tab:app-an}
\begin{tabular}{lm{1cm}|m{1cm}|m{1cm}|m{1cm}|m{1cm}|m{1cm}|m{1cm}|m{1cm}|}
\hline

\multicolumn{1}{|c|}{} & \multicolumn{1}{c|}{\cellcolor[gray]{0.9}} & \multicolumn{1}{c|}{\cellcolor[gray]{0.9}} & \multicolumn{1}{c|}{\cellcolor[gray]{0.9}} & \multicolumn{1}{c|}{\cellcolor[gray]{0.9}} & \multicolumn{1}{c|}{\cellcolor[gray]{0.9}} & \multicolumn{1}{c|}{\cellcolor[gray]{0.9}} & \multicolumn{1}{c|}{\cellcolor[gray]{0.9}}  & \multicolumn{1}{c|}{\cellcolor[gray]{0.7}}\\

\multicolumn{1}{|c|}{\multirow{-2}{*}{{ Potential Feature\textbackslash Category}}} & \multicolumn{1}{c|}{\multirow{-2}{*}{\cellcolor[gray]{0.9}Game}} & \multicolumn{1}{c|}{\multirow{-2}{*}{\cellcolor[gray]{0.9}Business}} & \multicolumn{1}{c|}{\multirow{-2}{*}{\cellcolor[gray]{0.9}Book}} & \multicolumn{1}{c|}{\multirow{-2}{*}{\cellcolor[gray]{0.9}Comics}} & \multicolumn{1}{c|}{\multirow{-2}{*}{\cellcolor[gray]{0.9}\begin{tabular}[c]{@{}c@{}}Commu-\\ nication\end{tabular}}} & \multicolumn{1}{c|}{\multirow{-2}{*}{\cellcolor[gray]{0.9}\begin{tabular}[c]{@{}c@{}}Edu-\\cation\end{tabular}}} & \multicolumn{1}{c|}{\multirow{-2}{*}{\cellcolor[gray]{0.9}\begin{tabular}[c]{@{}c@{}}Enterta-\\ inement\end{tabular}}}
& \multicolumn{1}{c|}{\multirow{-2}{*}{\cellcolor[gray]{0.7}\begin{tabular}[c]{@{}c@{}}Total \end{tabular}}}\\ \hline \hline

\multicolumn{9}{|c|}{{\cellcolor[gray]{0.6}Third-Party App Stores}} \\ \hline \hline

\multicolumn{1}{|l|}{\cellcolor[gray]{0.8}Stealthy Screenshots} & 
\multicolumn{1}{c|}{18 (90.00\%)} & \multicolumn{1}{c|}{11 (68.75\%)} & \multicolumn{1}{c|}{1 (100.00\%)} & \multicolumn{1}{c|}{2 (100.00\%)} &\multicolumn{1}{c|}{20 (95.24\%)}  & \multicolumn{1}{c|}{0 (0.00\%)} & \multicolumn{1}{c|}{9 (69.23\%)} & \multicolumn{1}{c|}{\cellcolor[gray]{0.7}61 (82.43\%)}  \\ \hline

\multicolumn{1}{|l|}{\cellcolor[gray]{0.8}Stealthy Photos} & 
\multicolumn{1}{c|}{0 (0.00\%)} & \multicolumn{1}{c|}{4 (25.00\%)} & \multicolumn{1}{c|}{0 (0.00\%)} & \multicolumn{1}{c|}{1 (50.00\%)} &\multicolumn{1}{c|}{13 (61.90\%)}  & \multicolumn{1}{c|}{0 (0.00\%)} & \multicolumn{1}{c|}{1 (7.69\%)} & \multicolumn{1}{c|}{\cellcolor[gray]{0.7}19 (25.68\%)} \\ \hline

\multicolumn{1}{|l|}{\cellcolor[gray]{0.8}Stealthy Videos} & 
\multicolumn{1}{c|}{0 (0.00\%)} & \multicolumn{1}{c|}{0 (0.00\%)} & \multicolumn{1}{c|}{0 (0.00\%)} & \multicolumn{1}{c|}{0 (0.00\%)} &\multicolumn{1}{c|}{12 (57.14\%)}  & \multicolumn{1}{c|}{0 (0.00\%)} & \multicolumn{1}{c|}{0 (0.00\%)} & \multicolumn{1}{c|}{\cellcolor[gray]{0.7}12 (16.22\%)} \\ \hline

\multicolumn{1}{|l|}{\cellcolor[gray]{0.8}Stealthy Audio} & 
\multicolumn{1}{c|}{2 (10.00\%)} & \multicolumn{1}{c|}{0 (0.00\%)} & \multicolumn{1}{c|}{0 (0.00\%)} & \multicolumn{1}{c|}{0 (0.00\%)} &\multicolumn{1}{c|}{12 (57.14\%)}  & \multicolumn{1}{c|}{0 (0.00\%)} & \multicolumn{1}{c|}{4 (30.77\%)} & \multicolumn{1}{c|}{\cellcolor[gray]{0.7}18 (24.32\%)} \\ \hline

\\[-0.15cm]

\cline{1-9}
 \multicolumn{1}{|c|}{\cellcolor[gray]{0.8}Number of Apps} & \multicolumn{1}{|c|}{20} &\multicolumn{1}{c|}{16}  & \multicolumn{1}{c|}{1} & \multicolumn{1}{c|}{2} & \multicolumn{1}{c|}{21} &\multicolumn{1}{c|}{1}  & \multicolumn{1}{c|}{13} & \multicolumn{1}{c|}{\cellcolor[gray]{0.7}74}  \\  
\cline{1-9}

\\[-0.15cm]

\hline

\multicolumn{9}{|c|}{{\cellcolor[gray]{0.6}Google Play Store}} \\ \hline \hline

\multicolumn{1}{|l|}{\cellcolor[gray]{0.8}Stealthy Screenshots} &
\multicolumn{1}{c|}{38 (95.00\%)} & \multicolumn{1}{c|}{33 (91.67\%)} & \multicolumn{1}{c|}{25 (83.33\%)} & \multicolumn{1}{c|}{30 (90.91\%)} &\multicolumn{1}{c|}{61 (87.14\%)}  & \multicolumn{1}{c|}{52 (67.53\%)} & \multicolumn{1}{c|}{37 (86.05\%)} & \multicolumn{1}{c|}{\cellcolor[gray]{0.7}276 (83.89\%)} \\ \hline

\multicolumn{1}{|l|}{\cellcolor[gray]{0.8}Stealthy Photos} &
\multicolumn{1}{c|}{3 (7.50\%)} & \multicolumn{1}{c|}{9 (25.00\%)} & \multicolumn{1}{c|}{2 (6.67\%)} & \multicolumn{1}{c|}{4 (12.12\%)} &\multicolumn{1}{c|}{32 (45.71\%)}  & \multicolumn{1}{c|}{4 (5.12\%)} & \multicolumn{1}{c|}{7 (16.28\%)} & \multicolumn{1}{c|}{\cellcolor[gray]{0.7}61 (18.54\%)} \\ \hline

\multicolumn{1}{|l|}{\cellcolor[gray]{0.8}Stealthy Videos} & 
\multicolumn{1}{c|}{0 (0.00\%)} & \multicolumn{1}{c|}{2 (5.56\%)} & \multicolumn{1}{c|}{1 (3.33\%)} & \multicolumn{1}{c|}{1 (3.03\%)} &\multicolumn{1}{c|}{27 (38.57\%)}  & \multicolumn{1}{c|}{1 (1.30\%)} & \multicolumn{1}{c|}{3 (6.98\%)} & \multicolumn{1}{c|}{\cellcolor[gray]{0.7}35 (10.64\%)} \\ \hline

\multicolumn{1}{|l|}{\cellcolor[gray]{0.8}Stealthy Audio} & 
\multicolumn{1}{c|}{0 (0.00\%)} & \multicolumn{1}{c|}{3 (8.33\%)} & \multicolumn{1}{c|}{2 (6.67\%)} & \multicolumn{1}{c|}{3 (9.09\%)} &\multicolumn{1}{c|}{30 (42.96\%)}  & \multicolumn{1}{c|}{6 (7.79\%)} & \multicolumn{1}{c|}{4 (9.30\%)} & \multicolumn{1}{c|}{\cellcolor[gray]{0.7}48 (14.59\%)} \\ \hline

\\[-0.15cm]

\cline{1-9}
\multicolumn{1}{|c|}{\cellcolor[gray]{0.8}Number of Apps} & \multicolumn{1}{|c|}{40} &\multicolumn{1}{c|}{36}  & \multicolumn{1}{c|}{30} & \multicolumn{1}{c|}{33} & \multicolumn{1}{c|}{70} &\multicolumn{1}{c|}{77}  & \multicolumn{1}{c|}{43} & \multicolumn{1}{c|}{\cellcolor[gray]{0.7}329}  \\  
\cline{1-9}

\\[-0.15cm]

\end{tabular}
\end{table*}

The analysis of 74 apps from third-party app stores \cite{MobileApkWorld,ApksFree} and 329 apps from the official Google Play \cite{GooglePlay}, shows concerning results, summarized in Table \ref{tab:app-an}. In particular, many of the analyzed apps could potentially behave as RAT, since they have the necessary permissions to perform stealthy operations targeting on-board I/O devices. For example, from the Google Play, 83.89\% of apps could potentially take stealthy screenshots. Furthermore, 25.68\% of apps from third-party app stores could potentially take stealthy photos. In each cell of Table \ref{tab:app-an}, the first value represents the number, the second value the percentage of apps, among all the app analyzed in the same category, that have the permissions required to perform the stealthy operation specified in the first column. We are not aware if these apps are actually misusing their permission, but we want to point out that it is possible for these apps to misuse their permissions to perform stealthy operations, and by statically analyzing the set of Android permissions used by apps, it is by no mean sufficient to distinguish between purely benign apps and malicious apps. 

\subsection{Anti-Malware Tools Detection Analysis}
\label{anti}

The analysis results relative to RAT detection by the 15 most popular anti-malware apps, available on Google Play \cite{GooglePlay} and used on smartphones by millions of user around the world, are summarized in Table \ref{tools}. The \textit{Installs} column indicates the number of installs performed by users on their smartphones. The \textit{Reviews} column specifies how many people gave a personal review and a score for the anti-malware app on Google Play. Finally, the \textit{Score} column reports the average score received by the app over a scale of 5 by user reviewing the app itself. 

\begin{table*}[t]
\centering
\caption{RAT Detection by the 15 Most Popular Android Anti-Malware Tools}

\label{tools}
\scriptsize

\tabcolsep=0.07cm
\vspace*{+2pt}

\begin{tabular}{l|c|c|c|c|c|c|c|c|} 
\cline{2-9} 
\multicolumn{1}{c|}{\multirow{8}{*}{\begin{tabular}[c]{@{}c@{}}\textit{Legend:} \\\\ \cmark \hspace{1 mm} Malware Detected \\ \xmark   \hspace{1 mm} Malware Undetected \\ $\circ$ Malware Detected as Privacy Violation \\ $\otimes$ Anti-malware crashed during Scan \end{tabular}}}& 
%  $\triangle$\hspace{1 mm} Prevention Claimed \\ 

\multicolumn{1}{c|}{\multirow{8}{*}{ \begin{turn}{90} \hspace{-2mm} \begin{tabular}{@{}c@{}}PlaceRaider\end{tabular} \end{turn} }} & 
\multicolumn{1}{c|}{\multirow{8}{*}{\begin{turn}{90}\hspace{-2mm} \begin{tabular}{@{}c@{}}Soundcomber \end{tabular} \end{turn}}}& 
\multicolumn{1}{c|}{\multirow{8}{*}{\begin{turn}{90}\hspace{-2mm} \begin{tabular}{@{}c@{}}StealthyStalker \end{tabular} \end{turn}}}&
\multicolumn{1}{c|}{\multirow{8}{*}{\begin{turn}{90}  \begin{tabular}{@{}c@{}}\hspace{-3mm} Dendroid\end{tabular}\end{turn} }}& 

%\multicolumn{1}{c|}{\multirow{8}{*}{\begin{turn}{90} \hspace{-2mm} \begin{tabular}{@{}c@{}}Krysanec (mSpy) $\ast$\end{tabular} \end{turn} }} &

\multicolumn{1}{c|}{\multirow{8}{*}{\begin{turn}{90} \hspace{-2mm} \begin{tabular}{@{}c@{}}Krysanec \end{tabular} \end{turn} }} &

\multicolumn{1}{c|}{\multirow{8}{*}{\begin{turn}{90}\hspace{-2mm} \begin{tabular}{@{}c@{}}Installs \end{tabular} \end{turn}}}&

\multicolumn{1}{c|}{\multirow{8}{*}{\begin{turn}{90}\hspace{-2mm} \begin{tabular}{@{}c@{}}Reviews \end{tabular} \end{turn}}} &

\multicolumn{1}{c|}{\multirow{8}{*}{\begin{turn}{90}\hspace{-2mm} \begin{tabular}{@{}c@{}}Score (Scale of 5) \end{tabular} \end{turn}}} \\

%\multicolumn{1}{c|}{\multirow{8}{*}{\begin{turn}{90}\hspace{-2mm} \begin{tabular}{@{}c@{}} Security Alerts \end{tabular} \end{turn}}}\\  

&& & & & & &&\\  
&& & & & & &&\\  
&& & & & & &&\\  
&& & & & & &&\\ 
&& & & & & &&\\  
&& & & & & &&\\ 
&& & & & & &&\\  \hline

% \tikz[baseline=(char.base)]{\node[shape=circle,draw,inner sep=1pt] (char) {1}}

\multicolumn{1}{|l|}{ \cellcolor[gray]{0.8}360 Security Antivirus Boost $\star$} 
& \xmark &\xmark & \xmark & \cellcolor[gray]{0.9}\cmark & \xmark & 100M-500M & 8,092,733 & 4.6\\ \hline

\multicolumn{1}{|l|}{ \cellcolor[gray]{0.8}AndroHelm AntiVirus}
& \xmark & \xmark & \xmark & \xmark & \xmark & 100K-500K & 5,383 & 4.1\\ \hline

\multicolumn{1}{|l|}{ \cellcolor[gray]{0.8}TrustGo Antivirus \& Mobile Security}
& \xmark & \xmark & \xmark & \cellcolor[gray]{0.9}\cmark & \cellcolor[gray]{0.9}\cmark & 10M-50M & 283,332 & 4.5\\ \hline

\multicolumn{1}{|l|}{ \cellcolor[gray]{0.8}AVAST Mobile Security \& Antivirus} 
& \cellcolor[gray]{0.9}\cmark & \xmark & \cellcolor[gray]{0.9}\cmark & \cellcolor[gray]{0.9}\cmark & \cellcolor[gray]{0.9}\cmark & 100M-500M & 3,481,194 & 4.5\\ \hline

\multicolumn{1}{|l|}{ \cellcolor[gray]{0.8}AVG AntiVirus Security Scan $\star$}  
& \xmark & \xmark & \xmark & \cellcolor[gray]{0.9}\cmark & \cellcolor[gray]{0.9}\cmark & 100M-500M & 4,079,893 & 4.4\\ \hline

\multicolumn{1}{|l|}{ \cellcolor[gray]{0.8}Bitdefender Mobile Security \& Antivirus} 
& \xmark & \xmark & \xmark & \cellcolor[gray]{0.9}\cmark & \cellcolor[gray]{0.9}\cmark & 1M-5M & 33,909 & 4.3\\ \hline

\multicolumn{1}{|l|}{ \cellcolor[gray]{0.8}Cheetah Mobile Security Antivirus \& AppLock  $\star$} 
& \xmark & \xmark & \xmark & \cellcolor[gray]{0.99}$\circ$ & \cellcolor[gray]{0.9}\cmark & 100M-500M & 13,258,188 & 4.7\\ \hline

\multicolumn{1}{|l|}{ \cellcolor[gray]{0.8}Dr Web Anti-virus} 
& \xmark & \xmark & \xmark & \cellcolor[gray]{0.9}\cmark & \cellcolor[gray]{0.9}\cmark & 50M-100M & 863,913 & 4.5\\ \hline

\multicolumn{1}{|l|}{ \cellcolor[gray]{0.8}ESET Mobile Security \& Antivirus} 
& \xmark & \xmark & \xmark & \cellcolor[gray]{0.9}\cmark & \cellcolor[gray]{0.9}\cmark & 5M-10M & 300,470 & 4.6\\ \hline

\multicolumn{1}{|l|}{ \cellcolor[gray]{0.8}Kaspersky Internet Security} 
& \xmark & \xmark & \xmark & \cellcolor[gray]{0.9}\cmark & \cellcolor[gray]{0.9}\cmark & 10M-50M & 1,019,526 & 4.6\\ \hline

\multicolumn{1}{|l|}{ \cellcolor[gray]{0.8}Lookout Security \& Antivirus} 
& \xmark & \xmark & \xmark & \cellcolor[gray]{0.9}\cmark & \cellcolor[gray]{0.9}\cmark & 100M-500M & 830,941 & 4.4\\ \hline

\multicolumn{1}{|l|}{ \cellcolor[gray]{0.8}Malwarebytes Anti-Malware} 
& \xmark & \xmark & \xmark &  \cellcolor[gray]{0.99}$\otimes$ & \cellcolor[gray]{0.9}\cmark & 1M-5M & 53,990 & 4.3\\ \hline

\multicolumn{1}{|l|}{ \cellcolor[gray]{0.8}McAfee Security \& Antivirus} 
& \xmark & \xmark & \xmark & \cellcolor[gray]{0.9}\cmark & \xmark & 10M-50M & 334,162 & 4.3\\ \hline

\multicolumn{1}{|l|}{ \cellcolor[gray]{0.8}Symantec Norton Security \& Antivirus} 
& \xmark & \xmark & \xmark & \cellcolor[gray]{0.9}\cmark & \cellcolor[gray]{0.9}\cmark & 10M-50M & 509,521 & 4.4 \\ \hline

\multicolumn{1}{|l|}{ \cellcolor[gray]{0.8}NQ Security Lab Antivirus Free-Mobile Security} 
& \xmark & \xmark &\xmark& \xmark& \xmark & 10M-50M & 497,640 & 4.3\\ \hline

\end{tabular}

\end{table*}

All the anti-malware apps have been updated with the most recent malware database before starting the scan, indeed 3 anti-malware apps (marked with $\star$  in Table \ref{tab:app-an}) have detected the Stagefright vulnerability \cite{ACentral} that would allow malicious code to send fraudulent MMS, only recently discovered. The analysis results are based on a first scan before malware installation, and a second scan during the execution of the malware, when the anti-malware has been kept actively scanning for 10 minutes. Subsequently, three consecutive scans have been performed, after malware installation. After the first scan, the anti-malware has been configured to actively keep scanning. We made sure to select full/deep scan from the scanning options. The three successive scans have been manually activated to force the anti-malware to rescan the entire system. At each new malware installation the smartphone has been flashed again with a clean copy of the OS and anti-malware software installation. 

The analysis revealed that most anti-malware tools are able to detect well-known RATs (e.g., Dendroid and Krysanec). We believe that this is due to the fact that well-known RATs have been classified and a signature has been generated and distributed on the Web. On the other hand, proof-of-concept RATs (e.g., PlaceRaider, SoundComber and StealthyStalker) are unknown and gone undetected by anti-malware tools even though they use similar techniques used by well-known RATs. Exceptionally, the AVAST Mobile Security Anti-Malware identifies some malice in both PlaceRaider and StealthyStalker. 

On the Android OS side, an interesting finding was that,  at install time, an alert was triggered to block the installation of Krysanec. At the second attempt, the Android OS asked the user if to proceed with the installation anyhow. Additionally, while uninstalling the app, Krysanec attacked the operating system by exploiting privilege escalation.

\subsection{Static and Dynamic Analysis Tools}

%\subsubsection{Inefficacy of Static and Dynamic Analysis Tools}
\label{stealthystalker}

%We selected 2 static and 2 dynamic analysis tools  that by analyzing app \textit{apk} files are able to identify malicious code. For the analysis we have used \textit{.apk} files provided by research groups, or accidentally leaked on the Internet. 

%CopperDroid temporarily unavailable (expected October 15 as stated by the authors).
The analysis results relative to RAT detection by four state-of-the-art static and dynamic analysis tools are reported in Table \ref{tools}.

\textit{VirusTotal} \cite{VirusTotal}, originally developed by Hispasec and now own by Google, is a free service that analyzes suspicious files and URLs and facilitates the quick detection of viruses, worms, trojans, and all kinds of malware. It uses 56 different anti-malware products and 61 online scan engines to check for viruses. VirusTotal was selected by PC World as one of the best 100 products of 2007. As shown in Table \ref{tools}, VirusTotal detects well-know RATs (e.g., Dendroid and Krysanec). The two RATs are identified as malicious with a score of respectively 22/56 and 20/56. The nominator in the score fractions refers to the number of tools that identify the app as potentially malicious, the denominator indicates the total number of tools that have analyzed the app.

\textit{MassVet} \cite{chen2015finding} compares a submitted app with apps already on a market, focusing on the difference between those sharing a similar UI structure (indicating a possible repackaging relation), and the commonality among those seemingly unrelated. MassVet uses a “DiffCom” analysis on top of an efficient similarity comparison algorithm, which maps features of an app’s UI structure or a method’s control-flow graph to a value for a fast comparison. As shown in Table \ref{tools}, MassVet detects malicious code in 3 of the 5 RATs analyzed.

\begin{table}[]
\centering
\scriptsize
\setlength{\tabcolsep}{.5em} % REDUCE HORIZONTAL PADDING

\caption{RAT Detection via Static and Dynamic Analysis}
\label{tool}
\begin{tabular}{l|c|c|c|c|}
\cline{2-4}
& \multicolumn{2}{c|}{\cellcolor[HTML]{9B9B9B}Static Analysis} & \multicolumn{2}{c|}{\cellcolor[HTML]{9B9B9B}Dynamic Analysis} \\ \cline{2-4} 
& {\cellcolor[HTML]{C0C0C0}VirusTotal}    & {\cellcolor[HTML]{C0C0C0}MassVet}    & {\cellcolor[HTML]{C0C0C0}\begin{tabular}[c]{@{}c@{}}Google\\ Bouncer\end{tabular}} & {\cellcolor[HTML]{C0C0C0}CopperDroid} \\ \hline
\multicolumn{1}{|l|}{\cellcolor[HTML]{C0C0C0}PlaceRaider}     &     \xmark            &     \xmark        &\cellcolor[gray]{0.99}N/A          & \cellcolor[gray]{0.99}N/A       \\ \hline
\multicolumn{1}{|l|}{\cellcolor[HTML]{C0C0C0}SoundComber}     &    \xmark             &     \cellcolor[gray]{0.9}\cmark       &        \cellcolor[gray]{0.99}N/A        & \cellcolor[gray]{0.99}N/A      \\ \hline
\multicolumn{1}{|l|}{\cellcolor[HTML]{C0C0C0}StealthyStalker} &    \xmark             &     \cellcolor[gray]{0.9}\cmark       &        \xmark        &      \cellcolor[gray]{0.99}N/A     \\ \hline
\multicolumn{1}{|l|}{\cellcolor[HTML]{C0C0C0}Dendroid}        &     \cellcolor[gray]{0.9}\cmark            &      \xmark       &     \cellcolor[gray]{0.99}N/A &\cellcolor[gray]{0.99}N/A \\ \hline
\multicolumn{1}{|l|}{\cellcolor[HTML]{C0C0C0}Krysanec}        &      \cellcolor[gray]{0.9}\cmark           &    \cellcolor[gray]{0.9}\cmark        &  \cellcolor[gray]{0.99}N/A & \cellcolor[gray]{0.99}N/A\\ \hline
\end{tabular}
\\
\vspace{5pt}
\textit{Legend:} \hspace{0.5mm}\cmark \hspace{0.5mm} Malware Detected \hspace{2mm} \xmark\hspace{0.5mm} Malware Undetected  \hspace{22mm} N/A \hspace{0.5mm} Data Not Available 
\end{table}

\begin{figure}[]
\centering
\includegraphics[width=80mm]{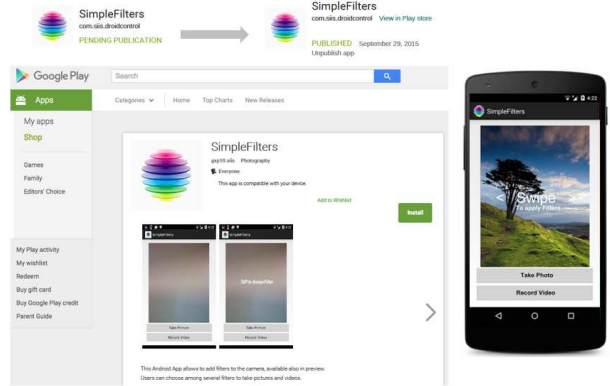}
\caption{StealthyStalker RAT app published on Google Play under the name of SimpleFilters}
\label{fig:bouncer}
\end{figure}

\textit{Google Bouncer} \cite{bouncer} is a codename used by Google, for a security service introduced early in 2012, to keep malicious apps off the official Google Play\footnote{According to Google, Bouncer was responsible for a 40\% drop in the number of malicious apps in its app store.}. Bouncer quietly and automatically scans apps (both new and previously uploaded ones) and developer accounts in Google Play with its reputation engine and cloud infrastructure. To test the effectiveness of Bouncer, in detecting RAT apps, we have implemented a proof-of-concept testing app, called \textit{StealthyStalker}\footnote{Submitted to Google Play under the name of \textit{SimpleFilters}}, able to take stealthy photos, videos, screenshots, record audio and hijack user-initiated operations.  To release an app through Google Play, a third-party developer has to participate in Android developer program and submit to Google for review. The app is signed and published by Google only after it passes the review process. As shown in Table \ref{tools} and Figure \ref{fig:bouncer}, the StealthyStalker app (submitted for publication with the fake name of SimpleFilter) successfully passed the Google Play (Bouncer) review and published, after a couple of hours, despite the hidden malicious code.  This means that Google Bouncer did not find any potential harm in the app. Following the ethical hacking practice, we immediately removed the app from Google Play before any user could actually download it, as proved by the download statistic provided by Google. Results for the other 4 RATs are not available due to the fact that we are not authorized to submit code written by other researchers or malicious developers.

\textit{CopperDroid} \cite{tam2015copperdroid} is a tool to perform dynamic analysis over Android apps to characterize the behavior of Android malware. It automatically analyzes low-level OS-specific and high-lelvel Android-specific behaviors of Android malware by observing and analyzing system call invocations, including IPC and RPC interactions, carried out as system calls.  Although CopperDroid is a powerful tool that dynamically analyzes Android apps, analysis results do not provide any hint on the maliciousness of a given sample. Indeed, by manually analyzing the logs files generated by CopperDroid (e.g, syscalls, pcap, logcat and basicbinder), we were not able to identify evidence of malice for the analyzed software.

\onecolumn

\subsection{Summary of selected Results from our User Study and Aware Evaluation}
\label{res}

%\subsection{Another appendix Subsection}

\begin{table*}[h]
\centering
\scriptsize

\setlength{\tabcolsep}{.2em} % REDUCE HORIZONTAL PADDING
\begin{tabular}{lllllllll}
\cline{1-3} \cline{5-6} \cline{8-9}
\multicolumn{3}{|l|}{\cellcolor[HTML]{9B9B9B}Demographic}                                            & \multicolumn{1}{l|}{} & \multicolumn{2}{l|}{\cellcolor[HTML]{9B9B9B}Mobile Platform Usage}                                 & \multicolumn{1}{l|}{} & \multicolumn{2}{l|}{\cellcolor[HTML]{9B9B9B}Participants Concerned about Personal Privacy}                   \\ \cline{1-3} \cline{5-6} \cline{8-9} 
\multicolumn{2}{|l|}{Total Participants}                                 & \multicolumn{1}{l|}{74}   & \multicolumn{1}{l|}{} & \multicolumn{1}{l|}{Smartphones}                                       & \multicolumn{1}{l|}{99\%} & \multicolumn{1}{l|}{} & \multicolumn{1}{l|}{When using Mobile Platfotm}                                 & \multicolumn{1}{l|}{43\%}  \\ \cline{1-3} \cline{5-6} \cline{8-9} 
\multicolumn{1}{|l|}{}                      & \multicolumn{1}{l|}{18-19} & \multicolumn{1}{l|}{6\%}  & \multicolumn{1}{l|}{} & \multicolumn{1}{l|}{Tablets}                                           & \multicolumn{1}{l|}{54\%} & \multicolumn{1}{l|}{} & \multicolumn{1}{l|}{When using third-party Apps}                                & \multicolumn{1}{l|}{57\%}  \\ \cline{2-3} \cline{5-6} \cline{8-9} 
\multicolumn{1}{|l|}{}                      & \multicolumn{1}{l|}{20-29} & \multicolumn{1}{l|}{76\%} & \multicolumn{1}{l|}{} & \multicolumn{1}{l|}{Wearables}                                         & \multicolumn{1}{l|}{10\%} & \multicolumn{1}{l|}{} & \multicolumn{1}{l|}{When using Platform's Camera}                               & \multicolumn{1}{l|}{62\%}  \\ \cline{2-3} \cline{5-6} \cline{8-9} 
\multicolumn{1}{|l|}{\multirow{-3}{*}{Age}} & \multicolumn{1}{l|}{30-39} & \multicolumn{1}{l|}{18\%} & \multicolumn{1}{l|}{} & \multicolumn{1}{l|}{Android OS}                                        & \multicolumn{1}{l|}{57\%} & \multicolumn{1}{l|}{} & \multicolumn{1}{l|}{When using Platform's Microphone}                           & \multicolumn{1}{l|}{55\%}  \\ \cline{1-3} \cline{5-6} \cline{8-9} 
\multicolumn{2}{|l|}{International Students}                             & \multicolumn{1}{l|}{70\%} & \multicolumn{1}{l|}{} & \multicolumn{1}{l|}{Apple iOS}                                         & \multicolumn{1}{l|}{42\%} &                       &                                                                                 &                            \\ \cline{1-3} \cline{5-6} \cline{8-9} 
\multicolumn{2}{|l|}{Nationalities}                                      & \multicolumn{1}{l|}{25}   & \multicolumn{1}{l|}{} & \multicolumn{1}{l|}{Other OS}                                          & \multicolumn{1}{l|}{1\%}  & \multicolumn{1}{l|}{} & \multicolumn{2}{l|}{\cellcolor[HTML]{9B9B9B}Participants' Awareness about RAT Operations}                    \\ \cline{1-3} \cline{5-6} \cline{8-9} 
\multicolumn{2}{|l|}{Non Computer Science Major's}                       & \multicolumn{1}{l|}{75\%} &                       &                                                                        &                           & \multicolumn{1}{l|}{} & \multicolumn{1}{l|}{Stealthy accesses to Platform's Camera}                     & \multicolumn{1}{l|}{56\%}  \\ \cline{1-3} \cline{5-6} \cline{8-9} 
\multicolumn{2}{|l|}{Major/Minor's Degrees}                              & \multicolumn{1}{l|}{34}   & \multicolumn{1}{l|}{} & \multicolumn{2}{l|}{\cellcolor[HTML]{9B9B9B}Security Behavior}                                     & \multicolumn{1}{l|}{} & \multicolumn{1}{l|}{Stealthy accesses to Platform's Microphone}                 & \multicolumn{1}{l|}{56\%}  \\ \cline{1-3} \cline{5-6} \cline{8-9} 
\multicolumn{3}{l}{}                                                                                 & \multicolumn{1}{l|}{} & \multicolumn{1}{l|}{Avoid downloading apps from unofficial App Stores} & \multicolumn{1}{l|}{78\%} & \multicolumn{1}{l|}{} & \multicolumn{1}{l|}{Knew about Dendroid or SoundComber}                         & \multicolumn{1}{l|}{4\%}   \\ \cline{1-3} \cline{5-6} \cline{8-9} 
\multicolumn{3}{|l|}{\cellcolor[HTML]{9B9B9B}Threats Identification without Aware}                   & \multicolumn{1}{l|}{} & \multicolumn{1}{l|}{Use a mobile anti-malware product}                 & \multicolumn{1}{l|}{24\%} & \multicolumn{1}{l|}{} & \multicolumn{1}{l|}{Knew about Krysanec or PlaceRaider}                         & \multicolumn{1}{l|}{3\%}   \\ \cline{1-3} \cline{5-6} \cline{8-9} 
\multicolumn{2}{|l|}{Participants identifying Attack A1}                 & \multicolumn{1}{l|}{36\%} &                       &                                                                        &                           &                       &                                                                                 &                            \\ \cline{1-3} \cline{5-6} \cline{8-9} 
\multicolumn{2}{|l|}{Participants identifying Attack A2}                 & \multicolumn{1}{l|}{0\%}  & \multicolumn{1}{l|}{} & \multicolumn{2}{l|}{\cellcolor[HTML]{9B9B9B}Threats Identification with Aware}                     & \multicolumn{1}{l|}{} & \multicolumn{2}{l|}{\cellcolor[HTML]{9B9B9B}Aware Effectiveness and Usability}                               \\ \cline{1-3} \cline{5-6} \cline{8-9} 
\multicolumn{2}{|l|}{Participants identifying Attack A3}                 & \multicolumn{1}{l|}{1\%}  & \multicolumn{1}{l|}{} & \multicolumn{1}{l|}{Participants identifying Attack A1}                & \multicolumn{1}{l|}{82\%} & \multicolumn{1}{l|}{} & \multicolumn{1}{l|}{Stealthy Accesses Blocked}                                  & \multicolumn{1}{l|}{100\%} \\ \cline{1-3} \cline{5-6} \cline{8-9} 
\multicolumn{2}{|l|}{Participants identifying Attack A4}                 & \multicolumn{1}{l|}{0\%}  & \multicolumn{1}{l|}{} & \multicolumn{1}{l|}{Participants identifying Attack A2}                & \multicolumn{1}{l|}{76\%} & \multicolumn{1}{l|}{} & \multicolumn{1}{l|}{User-Initiated Operations Hijacks Prevented}                & \multicolumn{1}{l|}{82\%}  \\ \cline{1-3} \cline{5-6} \cline{8-9} 
                                            &                            &                           &                       &                                                                        &                           & \multicolumn{1}{l|}{} & \multicolumn{1}{l|}{Rated Aware more secure than Original OS}    & \multicolumn{1}{l|}{90\%}  \\ \cline{8-9} 
                                            &                            &                           &                       &                                                                        &                           & \multicolumn{1}{l|}{} & \multicolumn{1}{l|}{Rated Aware as usable as the Original OS}    & \multicolumn{1}{l|}{80\%}  \\ \cline{8-9} 
                                            &                            &                           &                       &                                                                        &                           & \multicolumn{1}{l|}{} & \multicolumn{1}{l|}{Prefer Aware Gesture to Access Prompts on Screen} & \multicolumn{1}{l|}{57\%}  \\ \cline{8-9} 
                                            &                            &                           &                       &                                                                        &                           &                       &                                                                                 &                           
\end{tabular}
\end{table*}

\begin{figure}[h]
\centering

\begin{minipage}{1.0\textwidth}
\centering
\includegraphics[scale=0.29]{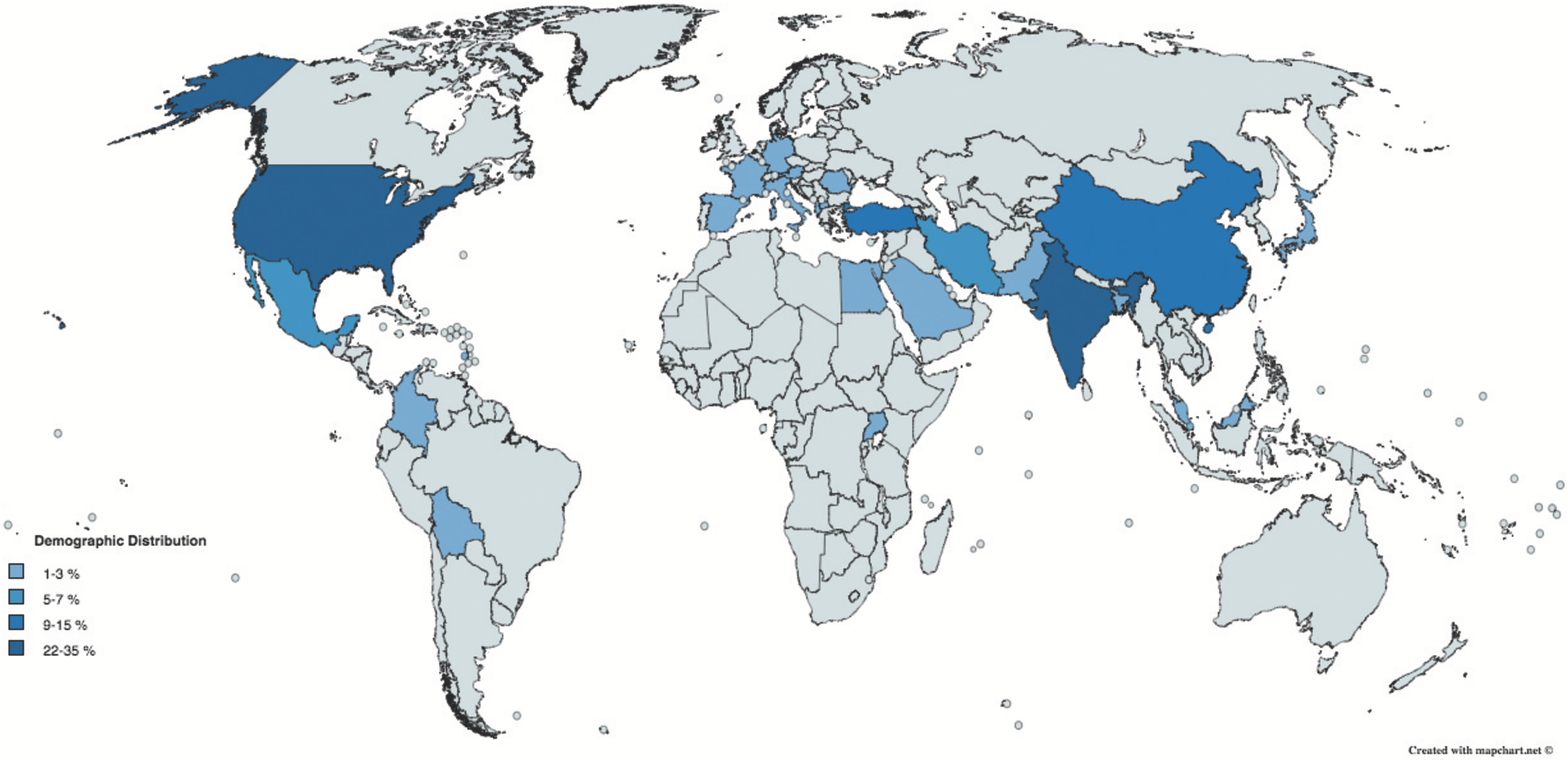}
\\Demographic Distribution of  Human Subjects participating in the Study \\
\end{minipage}
\begin{minipage}{1.0\textwidth}
\centering
\includegraphics[scale=0.25]{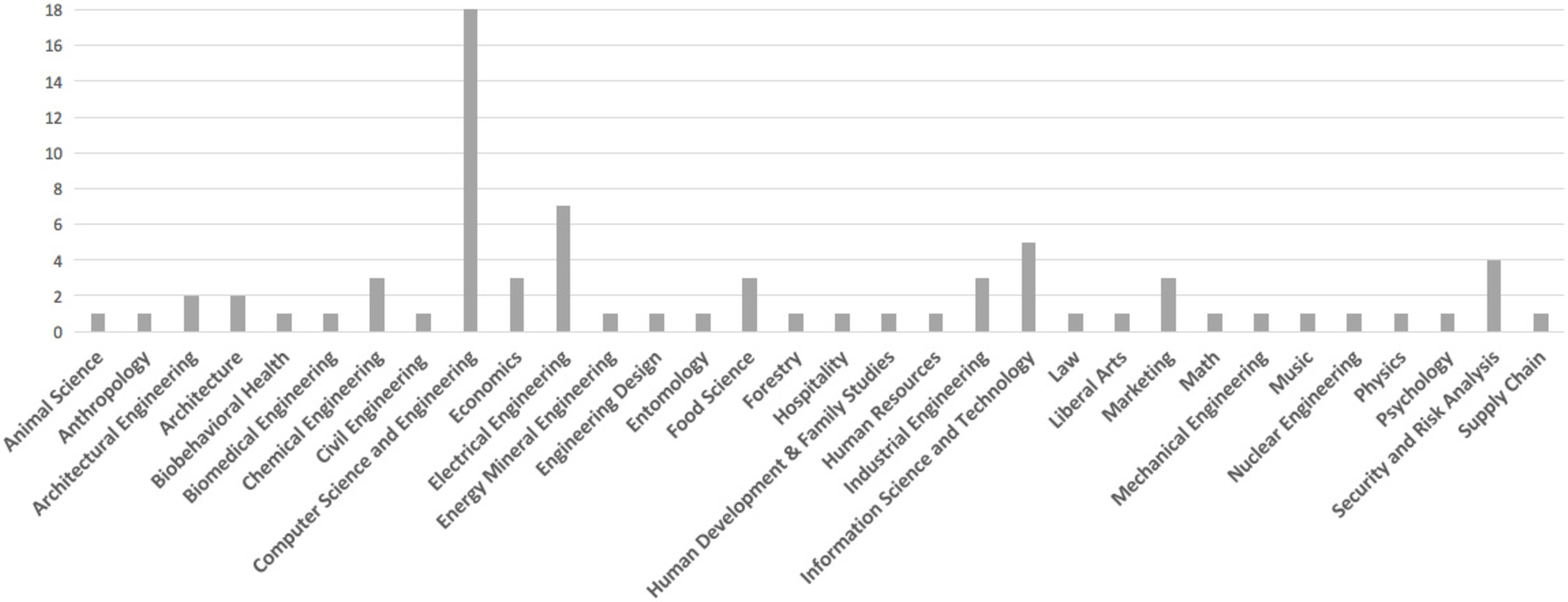}
\\ Major/Minor's Degree  Distribution of Human Subjects participating in the Study
\end{minipage}

\end{figure}

\end{document}

%% file: abstract.tex
{\bf Abstract } {\it Smartphones' cameras, microphones, and device displays enable users to capture and view memorable moments of their lives. However, adversaries can trick users into authorizing malicious apps that exploit weaknesses in current mobile platforms to misuse such on-board I/O devices to stealthily capture photos, videos, and screen content without the users' consent.
Contemporary mobile operating systems fail to prevent such misuse of I/O devices by authorized apps due to lack of binding between users' interactions and accesses to I/O devices performed by these apps. 
In this paper, we propose \system, a security framework for authorizing app requests to perform operations using I/O devices, which binds app requests with user intentions to make all uses of certain I/O devices explicit.
We evaluate our defense mechanisms through laboratory-based experimentation and a user study, involving 74 human subjects, whose ability to identify undesired operations targeting I/O devices increased significantly. Without \system, only 18\% of the participants were able to identify attacks from tested RAT apps. \system systematically blocks all the attacks in absence of user consent and supports users in identifying 82\% of social-engineering attacks tested to hijack approved requests, including some more sophisticated forms of social engineering not yet present in available RATs. \system introduces only 4.79\% maximum performance overhead over operations targeting I/O devices.  \system shows that a combination of system defenses and user interface can significantly strengthen defenses for controlling the use of on-board I/O devices.}

%% file: implementation.tex
\section{Implementation}

We have implemented a prototype of \system by modifying a recent release\footnote{A script to automatically integrate \system on top of previous versions of Android framework components and libraries is also available. \system source code will be made available on github.com.} of standard Android OS (version android-6.0.1\_r5) available through the Android Open Source Project (AOSP) \cite{aosp}. The \system prototype includes new components and modifies some pre-existing system services, libraries and the SystemUI app. Their footprint is about 325 LOC in C, 680 LOC in C++ and 882 LOC in Java. We tested the \system prototype on a Nexus 5 and Nexus 5X\footnote{Equipped with a fingerprint scanner.} smartphones. In the following paragraphs, we briefly describe the new and modified \system components.

 \textit{\system Hooks} are placed inside the original \textit{AudioSystem, MediaServer} and \textit{SensorManager} system services to capture the acquisition and release of I/O devices and the reading of sensor data. Other hooks are placed inside the original \textit{InputManager} and \textit{GestureDetector} to capture users' input events.
Hooks retrieve the PID of the calling processes, operations requested \textit{Opr} and target devices \textit{Dev}, which are then passed as a parameters in a call back to the \system Conditional Engine. 

\textit{\system Security Messages} are implemented by modifying the \textit{SurfaceManager, SurfaceFlinger, WindowManager} and the \textit{SystemUI}. In particular, an ImageView has been added to the Status Bar to display app IDs. Furthermore, a TextView has been added to display ongoing operations, and the service handling the status bar has been modified to receive and handle intent actions sent by the \system \textit{Conditional Engine}.

The \textit{\system Conditional Engine} is a new component added to the original Android framework. It includes: (1) the \textit{Callback Handler} in charge of processing callbacks from \system hooks, (2) the \textit{Verification Service}, in charge of validating preconditions, ongoing conditions, and exit conditions based on the information retrieved from callbacks by \system \textit{Hook}; and (3) the \textit{Conditional Rule Store}, designed to store and retrieve conditional rules used to enforce control over I/O devices.

The \textit{\system  Gesture Identification} module is implemented by modifying the \textit{InputManager, NativeInputManager, FingerprintManager, InputFlinger} and \textit{GestureDetector}, %Four new software modules are used to detect users' interaction with the Android OS and apps GUIs, which provide a stack structure receiving input from the lower level kernel drivers.  
to translate raw input data into higher-level events (key press, gesture, and such), and then propagates them to the \textit{Conditional Engine} through callbacks.

%% file: related_work.tex
\section{Related Work}

User-Driven Access Control (UDAC) \cite{roesner2012user} attempts to include users in the access control decision loop. Application level access control gadgets (ACGs) are used to verify that actions requested by apps genuinely comes from users. However, UDAC is subject to social-engineering attacks \cite{bianchi2015app,anderson2015supporting,soc_eng}, such as \textit{draw on top}\footnote{Drawing of graphical elements over other apps.} and \textit{app switch}\footnote{Malicious app replaces the legitimate top Activity with one of its own.}, because ACGs are displayed in a portion of the screen accessible to apps.

SemaDroid~\cite{xu2015semadroid} is a privacy-aware sensor management framework for smartphones that allows users to monitor sensors usage by installed apps, and control the disclosure of sensed data. However, users are supported in identifying undesired accesses only after apps have already performed operations targeting I/O device. 

Petracca \textit{et al.} \cite{audroid} propose AuDroid, an extension  to  the  SELinux  reference  monitor  integrated  into the Android OS, to enforce lattice policies over the dynamically changing use of system audio resources (e.g. microphone and speaker).  However, AuDroid only deals with accesses to microphone and speaker from third-party apps and does not provide a solid and general solution for other on-board I/O devices and sensors.

Bianchi \textit{et al.} \cite{bianchi2015app} address the problem of malicious apps surreptitiously replacing or mimicking the GUI of other apps to mount phishing and click-jacking attacks. However, Bianchi's solution cannot systematically prevent operations programmatically initiated by processes running apps, without users' interaction, because it does not provide the necessary mechanisms to bind users' interaction to access requests for I/O devices from processes running apps. Moreover, Bianchi's use visibility to transfer more responsibility to users when attack scenarios have to be identified, whereas visibility should be used to support users in making decision only when it is not possible to prevent attacks systematically.

%% file: arXiv16.bbl
\begin{thebibliography}{1}

\footnotesize
\vspace{-0.08in}

\bibitem {rogers}
 Dendroid malware can take over your camera, record audio, and sneak into Google Play. 
 https://blog.lookout.com/blog/ 2014/03/06/dendroid/

\vspace{-0.08in}
\bibitem {market}
  Smartphone {OS} Market Share. 
  http://www.idc.com/prodserv/ smartphone-os-market-share.jsp
\vspace{-0.08in}
 
\bibitem {lipovsky}
  Krysanec Trojan: Android backdoor lurking inside legitimate apps. 
  http://www.welivesecurity.com/ 2014/08/12/krysanec-trojan-android
\vspace{-0.08in}

\bibitem {AndroidDoc}
 Official Android Documentation. 
  \newblock http://developer.android.com
\vspace{-0.08in}

\bibitem {AndroidEnh}
  SEAndroid. 
  http://seandroid.bitbucket.org
\vspace{-0.08in}

\bibitem {aosp}
  Android Open Source Project.
  https://source.android.com/
\vspace{-0.08in}

\bibitem {JNI}
  Java Native Interface. 
  http://en.wikipedia.org/wiki/Java Native Interface
\vspace{-0.08in}


\bibitem {GooglePlay}
  Google Play app store. 
  https://play.google.com/
\vspace{-0.08in}

\bibitem {ACentral}
  Android Central. 
  http://www.androidcentral.com/stagefright
\vspace{-0.08in}

\bibitem {ApksFree}
  Apks Free app store. http://www.androidapksfree.com
\vspace{-0.08in}

\bibitem {MobileApkWorld}
  Mobile Apk World app store.
  http://mobileapkworld.com
\vspace{-0.08in}


\bibitem {VirusTotal}
  VirusTotal. Free Online Virus, Malware and URL Scanner. 
  https://www.virustotal.com
\vspace{-0.08in}

\bibitem {npr}
 Smartphones Are Used To Stalk, Control Domestic Abuse Victims. 
  http://www.npr.org/sections/alltechconsidered/2014/09/15/ 346149979/smartphones-are-used-to-stalk-control-domestic-abuse-victims
\vspace{-0.08in}

\bibitem {bouncer}
  Google Bouncer.
  http://blog.trendmicro.com/trendlabs-security- intelligence/a-look-at-google-bouncer/
\vspace{-0.08in}

\bibitem {esurance}
  Speed up your car insurance claim.
  https://www.esurance.com/ photo-claims
\vspace{-0.08in}

\bibitem {soc_eng}
  Social-Engineering Attacks.
  https://en.wikipedia.org/wiki/Social \_engineering\_(security)
\vspace{-0.08in}

\bibitem {pnc}
  PNC Mobile Banking.
  https://www.pnc.com/en/personal-banking/banking/online-and-mobile-banking/mobile-banking.html
\vspace{-0.08in}

\bibitem {knox}
  Samsung Know White Papers.
   https://www.samsungknox.com/ en/support/knox-workspace/white-papers
\vspace{-0.08in}


\bibitem {yee2004aligning}
  Yee, Ka-Ping.
  Aligning security and usability.
\vspace{-0.08in}

\bibitem {park2004ucon}
  Park, Jaehong and Sandhu, Ravi.
  The UCON ABC usage control model.
\vspace{-0.08in}

\bibitem {monitor}
  Anderson, James P.
  Computer Security Technology Planning Study. Volume 2.
\vspace{-0.08in}

\bibitem {templeman}
 Robert Templeman and Zahid Rahman and David Crandall and Apu Kapadia.
 PlaceRaider: Virtual theft in physical spaces with smartphones.
\vspace{-0.08in}

\bibitem {schlegel}
  Schlegel, Roman and Zhang, Kehuan and Zhou, Xiao-yong and Intwala, Mehool and Kapadia, Apu and Wang, XiaoFeng.
  Soundcomber: A Stealthy and Context-Aware Sound Trojan for Smartphones.
\vspace{-0.08in}

\bibitem {felt2012}
  Felt, Adrienne Porter and Ha, Elizabeth and Egelman, Serge and Haney, Ariel and Chin, Erika and Wagner, David.
  Android permissions: {U}ser attention, comprehension, and behavior.
\vspace{-0.08in}

\bibitem {roesner2012user}
  Roesner, Franziska and Kohno, Tohru and Moshchuk, Alexander and Parno, Bryan and Wang, Harry Jiannan and Cowan, Crispin.
  User-driven access control: {R}ethinking permission granting in modern operating systems.
\vspace{-0.08in}


\bibitem {roesner2014world}
  World-driven access control for continuous sensing
  Roesner, Franziska and Molnar, David and Moshchuk, Alexander and Kohno, Tadayoshi and Wang, Helen.
\vspace{-0.08in}

\bibitem {anderson2015supporting}
   Anderson, Fraser and Grossman, Tovi and Wigdor, Daniel and Fitzmaurice, George.
   Supporting Subtlety with Deceptive Devices and Illusory Interactions.
\vspace{-0.08in}

\bibitem {bianchi2015app}
  Bianchi, Antonio and Corbetta, Jacopo and Invernizzi, Luca and Fratantonio, Yanick and Kruegel, Christopher and Vigna, Giovanni.
  What the App is That? Deception and Countermeasures in the Android User Interface.
\vspace{-0.08in}

\bibitem {smalley2013security}
  Smalley, Stephen and Craig, Robert.
  Security Enhanced Android: Bringing Flexible MAC to Android.
  \vspace{-0.08in}

\bibitem {nadkarni2014asm}
  Nadkarni, Adwait and Enck, William.
  ASM: A programmable interface for extending Android security.
  \vspace{-0.08in}

\bibitem {backes2014android}
  Backes, Michael and Bugiel, Sven and Gerling, Sebastian and von Styp-Rekowsky, Philipp.
  Android Security Framework: Extensible multi-layered access control on Android.
  \vspace{-0.08in}

%\bibitem {horsch2014trustid}
%  Horsch, Julian and B{\"o}ttinger, Konstantin and Wei{\ss}, Michael and Wessel, Sascha and Stumpf, Frederic.
%  TrustID: {T}rustworthy identities for untrusted mobile devices.

\bibitem {chen2015finding}
  Chen, Kai and Wang, Peng and Lee, Yeonjoon and Wang, XiaoFeng and Zhang, Nan and Huang, Heqing and Zou, Wei and Liu, Peng.
  Finding unknown malice in 10 seconds: {M}ass vetting for new threats at the Google-Play scale.
  \vspace{-0.08in}


\bibitem {tam2015copperdroid}
  Tam, Kimberly and Khan, Salahuddin and Fattori, Aristide and Cavallaro, Lorenzo.
  CopperDroid: Automatic Reconstruction of Android Malware Behaviors.
  \vspace{-0.08in}


\bibitem {xu2015semadroid}
  Xu, Zhi and Zhu, Sencun.
  SemaDroid: {A} Privacy-Aware Sensor Management Framework for Smartphones.
  \vspace{-0.08in}

\bibitem {audroid}
  Petracca, Giuseppe and Sun, Yuqiong and Atamli, Ahmad and Jaeger, Trent.
  AuDroid: Preventing Attacks on Audio Channels in Mobile Devices.
  \vspace{-0.08in}
\
\bibitem {primal}
Primal Wijesekera and Arjun Baokar and Ashkan Hosseini and Serge Egelman and David Wagner and Konstantin Beznosov.
Android Permissions Remystified: {A} Field Study on Contextual Integrity.
\vspace{-0.08in}

\bibitem {Sheng10}
 Sheng, Steve and Holbrook, Mandy and Kumaraguru, Ponnurangam and Cranor, Lorrie Faith and Downs, Julie.
 Who Falls for Phish? A Demographic Analysis of Phishing Susceptibility and Effectiveness of Interventions.
 \vspace{-0.08in}


\bibitem {Bohme11}
 B\"{o}hme, Rainer and Grossklags, Jens.
 The Security Cost of Cheap User Interaction.
 \vspace{-0.08in}







\end{thebibliography}
